\documentclass[prd, 11pt, aps, hyphen, nofootinbib, preprintnumbers,
showpacs, superscriptaddress, tightenlines]{revtex4}

\usepackage{amsmath}
\usepackage{amssymb}
\usepackage[english]{babel}
\usepackage{color}
\usepackage{dcolumn}
\usepackage{epsfig}
\usepackage{epstopdf}
\usepackage{graphicx}
\usepackage[latin9]{inputenc}
\usepackage{subfigure}
\usepackage{url}

\usepackage[bookmarks, bookmarksopen, bookmarksnumbered]{hyperref}
\usepackage[all]{hypcap}
\newcommand{\DOT}{.}

\newcommand{\be}{\begin{equation}}
\newcommand{\ee}{\end{equation}}

\begin{document}

\title{Turduckening black holes: an analytical and computational study}

\author{David~Brown}

\affiliation{Department of Physics, North Carolina State University,
  Raleigh, NC 27695, USA}

\author{Peter~Diener}

\affiliation{Center for Computation \& Technology,
  Louisiana State University, Baton Rouge, LA 70803, USA}
\homepage{http://www.cct.lsu.edu/}

\affiliation{Department of Physics \& Astronomy,
  Louisiana State University, Baton Rouge, LA 70803, USA}
\homepage{http://relativity.phys.lsu.edu/}

\author{Olivier~Sarbach}

\affiliation{Instituto de F\a'{\i}sica y Matem\a'aticas,
Universidad Michoacana de San Nicol\a'as de Hidalgo, Edificio C-3,
Cd.\ Universitaria, C. P. 58040 Morelia, Michoac\a'an, M\a'exico}

\author{Erik~Schnetter}

\affiliation{Center for Computation \& Technology,
  Louisiana State University, Baton Rouge, LA 70803, USA}
\homepage{http://www.cct.lsu.edu/}

\affiliation{Department of Physics \& Astronomy,
  Louisiana State University, Baton Rouge, LA 70803, USA}
\homepage{http://relativity.phys.lsu.edu/}

\author{Manuel~Tiglio}

\affiliation{Center for Fundamental Physics, Department of Physics, University of Maryland, College Park, MD
  20742, USA}

\affiliation{Center for Scientific Computation and Mathematical Modeling, University of Maryland, College Park, MD
  20742, USA}

\begin{abstract}

We provide a detailed analysis of several aspects of the turduckening
technique for evolving black holes. At the analytical level we study
the constraint propagation for a general family of BSSN-type formulation of Einstein's field equations 
and identify under what conditions the turducken procedure is
rigorously justified and under what conditions constraint violations
will propagate to the outside of the black holes. We present
high-resolution spherically symmetric studies which verify our
analytical predictions. Then we present three-dimensional simulations
of single distorted black holes using different variations of the
turduckening method and also the puncture method. We study the effect
that these different methods have on the coordinate conditions,
constraint violations, and extracted gravitational waves.
We find that the waves agree up to small but non-vanishing
differences, caused by escaping superluminal gauge modes.  These
differences become smaller with increasing detector location.

\end{abstract}


\maketitle

%%%%%%%%%%%%%%%%%%%%%%%%%%%%%%%%%%%%%%%%%%%%%%%%%%%%%%%%%%%%%%%%%%%%%%%%%%%%%
\section{Introduction}
%%%%%%%%%%%%%%%%%%%%%%%%%%%%%%%%%%%%%%%%%%%%%%%%%%%%%%%%%%%%%%%%%%%%%%%%%%%%%

In a previous publication \cite{Brown2007b} we discussed the
\emph{turduckening} approach to numerical simulations of black holes
in Einstein's theory.  The technique relies on initially smoothing the
data inside each black hole and solving the Einstein evolution
equations everywhere at later times. The idea was first proposed in
Ref.~\cite{Bona04a} under the name of ``free black hole evolution''.
It shares many similarities with the ``stuffed black hole''
\cite{Arbona97,Arbona99} and ``magic matter''~\cite{Misner01}
approaches.  In Ref.~\cite{Brown2007b} we presented a particular
implementation that works in practice for binary black holes. We also
provided justification for our implementation, and numerical evidence
of the geometrical picture behind it. Complementary results based on
very similar ideas were independently found and presented in
Ref.~\cite{Etienne2007a} under the name ``filling the holes''.

The intuitive rationale behind the turduckening approach is that the
physics in the exterior of a black hole should be causally
disconnected from the unphysical smoothing in the interior. This is
the same rationale behind black hole excision
\cite{Unruh84,Thornburg87}, but here one proceeds in a different
way. In particular, one does not need to place an inner boundary per
black hole in order to remove the interior. The computational domain
in this technique is trivial from a topological point of view, and
therefore the discretization remains simple. Thus, the method shares
the simplicity of the moving punctures technique
\cite{Campanelli:2005dd,Baker:2005vv} but is not restricted to
puncture--type initial data and does not require regularization of the
equations near special points.

In this paper we extend our analysis of the turducken technique,
concentrating on both conceptual and practical issues.

We begin in Section \ref{sec:formulation} by describing the
formulation of the equations that we used in Ref.~\cite{Brown2007b},
which is a specific version of the BSSN-type family. We analyze in
detail the hyperbolicity of both the main system and the subsidiary
(constraint) system, placing particular emphasis on the propagation
speeds of constraint violating modes.  It is well known that in the
Einstein equations the ``true'' degrees of freedom are coupled to
coordinate and constrained degrees of freedom. One therefore needs to
guarantee that, for the formulation of the Einstein evolution
equations and the gauge conditions being used, the smoothing in the
interior of each black hole does not affect the ``physics'' in the
exterior. This is a non-trivial condition and, in fact, it is
formulation and gauge dependent. In Section \ref{sec:formulation} we
show that there are some versions of the BSSN equations where this
condition does \emph{not} hold, and where constraint violations that
originate in the interior of the black hole \emph{do} propagate to the
outside.  However, we are also able to identify a class of BSSN--type
equations for which we can rigorously guarantee that constraint
violations inside the black hole do not leak to the outside.

Next we concentrate on the issue of whether gauge modes can escape
from the interior of the black hole. The gauge conditions that we use
are those of the moving punctures technique. In Section
\ref{sec:formulation} we show that some of the characteristic speeds
depend on the solution itself. Therefore it is not possible to
determine a priori whether or not some modes will become
superluminal. There is nothing wrong with modes leaking from the black
hole interior, as long as these modes represent the gauge freedom
inherent in the evolution problem.  It is nevertheless of conceptual
and practical importance to understand how the turduckening procedure
might affect the gauge outside the black hole. Below we turn to this
point by analyzing the numerical data.

Having analyzed the system of equations at the continuum, and in
particular, having shown that at that level the turduckening procedure
does not introduce constraint violations to the exterior of a black
hole, we address the discretization and numerical implementation in
the following sections. We begin in Sec.\ \ref{sec:code} with a brief
description of the numerical codes that are used in this paper.  In
Sec.~\ref{sec:KS} we evolve turduckened initial data for a
Schwarzschild black hole with a spherically symmetric one-dimensional
code. Using this code we can corroborate with high numerical accuracy
that the constraint violations in the formulation of the equations
that we use do not leak to the outside, as expected from our
analytical analysis.

The one-dimensional numerical studies also reveal an interesting
property of turducken evolutions: even though the stuffing procedure
initially introduces large constraint violations inside the black
holes, these violations quickly decay to very small values as the
evolution proceeds. This occurs because the shift vector quickly moves
the coordinate grid points away from the future domain of dependence
of the turduckened region, while
the constraint-violating modes are confined to the inside of the black
hole. The numerical data then relax to a portion
of the stationary $1+\log$ ``trumpet slice'' of the black hole
\cite{Hannam:2006vv1, Brown:2007, Hannam2008a}.
This is the same end state as obtained
with puncture evolution.

We also use one-dimensional simulations to investigate the
possibility of superluminal gauge modes. We find that gauge modes are
in fact superluminal and propagate from the interior to the exterior
of the black hole.  In particular, the smoothing procedure affects the
coordinate conditions outside the black hole. However, we find that
the differences in gauge that arise from different types of smoothing
quickly decay in time. As already mentioned, we find that the
turducken solution approaches a portion of the trumpet slice,
regardless of the type of smoothing.

Given that it has already been shown in \cite{Brown2007b,Etienne2007a}
that the turduckening procedure works in practice for binary black
hole evolutions, we next analyze in detail several aspects of single
black hole evolutions.  In Sec.~\ref{sec:BY} we present results from
three-dimensional evolutions of a single distorted rotating black
hole. This data is obtained by applying the smoothing procedure to
puncture initial data with Bowen--York extrinsic curvature. We compare
in detail the gauge conditions and extracted waveforms produced in
calculations with turduckening regions of different sizes as well as a
pure puncture evolution. We also show that we have fourth order
convergence in the extracted waveforms.

We find, in agreement with the one-dimensional results, that
superluminal gauge modes are able to propagate to the outside of the
horizon.  However, if the turduckening region is sufficiently small,
the effect of these gauge modes decreases with radius outside of the
black hole, and becomes small enough that for practical purposes it
can be disregarded in 3D simulations.

Comparing the waveforms from turduckening and pure puncture runs, we
find that the differences are very small and that most of them
converge to zero with increasing resolution. The remaining differences
are caused by the differences in gauge at the finite detector
locations and are found to become smaller with increasing detector
location.

Section \ref{sec:Final} contains some final remarks. 

%%%%%%%%%%%%%%%%%%%%%%%%%%%%%%%%%%%%%%%%%%%%%%%%%%%%%%%%%%%%%%%%%%%%%%%%%%%%%
\section{Formulation of the equations, constraint propagation, hyperbolicity,
and characteristic speeds}
\label{sec:formulation}
%%%%%%%%%%%%%%%%%%%%%%%%%%%%%%%%%%%%%%%%%%%%%%%%%%%%%%%%%%%%%%%%%%%%%%%%%%%%%

In this section we first give the explicit form of the evolution and
constraint equations used in our Cauchy formulation of Einstein's
field equations. It is a special case of the family of formulations
analyzed in \cite{Beyer:2004sv}. Next, we summarize the conditions
under which this formulation is hyperbolic and give the characteristic
speeds. Finally, we extend the analysis performed in \cite{Beyer:2004sv} by
deriving the constraint propagation system, describing the propagation
of constraint violations, and analyzing its hyperbolic structure. In
particular, we give necessary conditions for this system to be
symmetric hyperbolic and possess no superluminal speeds. We then prove
that under these conditions constraint violations {\em inside} a black
hole which are present in the turducken approach cannot propagate to
the domain of outer communication.

\subsection{Formulation of the equations}

As mentioned in the introduction, we consider a BSSN-type formulation
of Einstein's equations where the three metric $\gamma_{ij}$ and the
extrinsic curvature $K_{ij}$ are decomposed according to
\begin{eqnarray}
\gamma_{ij} &=& e^{4\phi}\tilde{\gamma}_{ij}\; ,\\
K_{ij} &=& e^{4\phi}\left( \tilde{A}_{ij} 
 + \frac{1}{3}\tilde{\gamma}_{ij} K \right) .
\end{eqnarray}
Here, the conformal factor $e^{2\phi}$ is chosen such the conformal
metric $\tilde{\gamma}_{ij}$ has unit determinant, and $K =
\gamma^{ij} K_{ij}$ and $\tilde{A}_{ij}$ are the trace and the
trace-less part, respectively, of the conformally rescaled extrinsic
curvature. The $3+1$ decomposition of Einstein's equations along with
suitable gauge conditions for lapse ($\alpha$) and shift ($\beta^i$)
yields the following evolution system
\cite{Beyer:2004sv}\footnote{There are two sign errors in Eqs.\ (5)
and (6) of Ref.\ \cite{Beyer:2004sv}. The first is in front of the
second term of Eq.\ (5) and the second in front of the fourth term in
Eq.\ (6). Since these errors only affect lower order terms they do not
affect the results in \cite{Beyer:2004sv} in any way. We thank Dae-Il
Choi for pointing out these errors to us.}
\begin{widetext}
\begin{eqnarray}
\hat{\partial}_0 \alpha &=& -\alpha^2 f(\alpha,\phi,x^\mu) (K - K_0(x^\mu)),
\label{Eq:BSSN1}\\
%%%%%%%%%%%%%%%%%%%
\hat{\partial}_0 K &=& -e^{-4\phi}\left[ 
 \tilde{D}^i\tilde{D}_i \alpha + 2\partial_i\phi \cdot\tilde{D}^i\alpha \right]
 + \alpha\left( \tilde{A}^{ij}\tilde{A}_{ij} + \frac{1}{3} K^2 \right)
 - \alpha S,
\label{Eq:BSSN2}\\
%%%%%%%%%%%%%%%%%%%
\hat{\partial}_0 \beta^i &=& \alpha^2 G(\alpha,\phi,x^\mu) B^i,
\label{Eq:BSSN3}\\
%%%%%%%%%%%%%%%%%%%
\hat{\partial}_0 B^i &=& e^{-4\phi} H(\alpha,\phi,x^\mu)
  \hat{\partial}_0\tilde{\Gamma}^i - \eta^i(B^i,\alpha,x^\mu)
\label{Eq:BSSN4}\\
%%%%%%%%%%%%%%%%%%%
\hat{\partial}_0 \phi &=& -\frac{\alpha}{6}\, K + \frac{1}{6}\partial_k\beta^k,
\label{Eq:BSSN5}\\
%%%%%%%%%%%%%%%%%%%
\hat{\partial}_0 \tilde{\gamma}_{ij} &=& -2\alpha\tilde{A}_{ij} 
 + 2\tilde{\gamma}_{k(i}\partial_{j)}\beta^k 
 - \frac{2}{3}\tilde{\gamma}_{ij}\partial_k\beta^k ,
\label{Eq:BSSN6}\\
%%%%%%%%%%%%%%%%%%%
\hat{\partial}_0 \tilde{A}_{ij} &=& e^{-4\phi}\left[ 
 \alpha\tilde{R}_{ij} + \alpha R^\phi_{ij} - \tilde{D}_i\tilde{D}_j\alpha 
  + 4\partial_{(i}\phi\cdot\tilde{D}_{j)}\alpha\right]^{TF}
\nonumber\\
 && {} + \alpha K\tilde{A}_{ij} - 2\alpha\tilde{A}_{ik}\tilde{A}^k_{\; j}
  + 2\tilde{A}_{k(i}\partial_{j)}\beta^k 
  - \frac{2}{3}\tilde{A}_{ij}\partial_k\beta^k
  - \alpha e^{-4\phi} \hat{S}_{ij} ,
\label{Eq:BSSN7}\\
%%%%%%%%%%%%%%%%%%%
\hat{\partial}_0\tilde{\Gamma}^i &=& 
 \tilde{\gamma}^{kl}\partial_k\partial_l\beta^i
 + \frac{1}{3} \tilde{\gamma}^{ij}\partial_j\partial_k\beta^k 
 + \partial_k\tilde{\gamma}^{kj} \cdot \partial_j\beta^i
 - \frac{2}{3}\partial_k\tilde{\gamma}^{ki} \cdot \partial_j\beta^j\nonumber\\
 && {} - 2\tilde{A}^{ij}\partial_j\alpha 
 + 2\alpha\left[ (m-1)\partial_k\tilde{A}^{ki} - \frac{2m}{3}\tilde{D}^i K
    + m(\tilde{\Gamma}^i_{\; kl}\tilde{A}^{kl} + 6\tilde{A}^{ij}\partial_j\phi)
\right] - S^i,
\label{Eq:BSSN8}
\end{eqnarray}
\end{widetext}
where we have introduced the operator $\hat{\partial}_0 = \partial_t -
\beta^j\partial_j$. Here, all quantities with a tilde refer to the
conformal three metric $\tilde{\gamma}_{ij}$ and the latter is used in
order to raise and lower indices. In particular, $\tilde{D}_i$
and $\tilde{\Gamma}^k_{\; ij}$ refer to the covariant derivative and
the Christoffel symbols, respectively, with respect to
$\tilde{\gamma}_{ij}$. The expression $[ \cdots ]^{TF}$ denotes the
trace-less part (with respect to the metric $\tilde{\gamma}_{ij}$) of
the expression inside the parentheses, and
\begin{widetext}
\begin{eqnarray}
\tilde{R}_{ij} 
 &=& -\frac{1}{2} \tilde{\gamma}^{kl}\partial_k\partial_l\tilde{\gamma}_{ij} 
  + \tilde{\gamma}_{k(i}\partial_{j)}\tilde{\Gamma}^k
  - \tilde{\Gamma}_{(ij)k}\partial_l\tilde{\gamma}^{lk} 
  + \tilde{\gamma}^{ls}\left( 2\tilde{\Gamma}^k_{\; l(i}\tilde{\Gamma}_{j)ks} 
  + \tilde{\Gamma}^k_{\; is}\tilde{\Gamma}_{klj} \right),
\\
R^\phi_{ij} &=& -2\tilde{D}_i\tilde{D}_j\phi 
  - 2\tilde{\gamma}_{ij} \tilde{D}^k\tilde{D}_k\phi
  + 4\tilde{D}_i\phi\, \tilde{D}_j\phi 
  - 4\tilde{\gamma}_{ij}\tilde{D}^k\phi\, \tilde{D}_k\phi.
\end{eqnarray}
\end{widetext}

The gauge conditions imposed on the lapse, Eq.\ (\ref{Eq:BSSN1}), is a
generalization of the Bona-Massó condition \cite{Bona95b} where
$f(\alpha,\phi,x^\mu)$ is a smooth and strictly positive function and
$K_0(x^\mu)$ is an arbitrary smooth function. The conditions imposed
on the shift in Eqs.\ (\ref{Eq:BSSN3},\ref{Eq:BSSN4}) is a
generalization of the hyperbolic Gamma driver \cite{Alcubierre02a}
condition where $G(\alpha,\phi,x^\mu)$ and $H(\alpha,\phi,x^\mu)$ are
smooth, strictly positive functions, and $\eta^i(B^j,\alpha,x^\mu)$ is
a smooth vector-valued function. The term
$\hat{\partial}_0\tilde{\Gamma}^i$ in Eq.\ (\ref{Eq:BSSN4}) is set
equal to the right-hand side of Eq.\ (\ref{Eq:BSSN8}). Note that we
use the operator $\hat{\partial}_0$ (as opposed to $\partial_t$) in
these gauge conditions; not only does this simplify the analysis of
the principal part of the evolution equations, it also results in
stable binary black hole evolutions for moving punctures
\cite{vanMeter:2006vi} and the turducken approach \cite{Brown2007b}.

Finally, the parameter $m$ which was introduced in
\cite{Alcubierre99e}, controls how the momentum constraint is added to
the evolution equations for the variable $\tilde{\Gamma}^i$. The
standard choice in numerical simulations is $m=1$ which eliminates the
divergence of $\tilde{A}^{ij}$ in Eq.\ (\ref{Eq:BSSN8}). However, we
find it instructive not to fix $m=1$ in this article. The source terms
$S$, $\hat{S}_{ij}$ and $S^i$ are defined in terms of the four Ricci
tensor, $R^{(4)}_{ij}$, and the constraint variables
\begin{eqnarray}
H &\equiv& 
 \frac{1}{2}\left( \gamma^{ij} R^{(\gamma)}_{ij} + K^2 - K^{ij} K_{ij} \right),
\label{Eq:BSSNCons1}\\
M_i &\equiv& \tilde{D}^j \tilde{A}_{ij} - \frac{2}{3} \tilde{D}_i K 
 + 6\tilde{A}_{ij} \tilde{D}^j\phi,
\label{Eq:BSSNCons2}\\
C^i_\Gamma &\equiv& \tilde{\Gamma}^i + \partial_j\tilde{\gamma}^{ij}\; ,
\label{Eq:BSSNCons3}
\end{eqnarray}
as
\begin{eqnarray}
S &=& \gamma^{ij} R^{(4)}_{ij} - 2 H,
\label{Eq:DefS}\\
\hat{S}_{ij} &=& \left[ R^{(4)}_{ij} 
 + \tilde{\gamma}_{k(i}\partial_{j)} C_\Gamma^k \right]^{TF},
\label{Eq:DefSij}\\
S^i &=& 2\alpha\, m\,\tilde{\gamma}^{ij} M_j - \hat{\partial}_0 C_\Gamma^i\; .
\label{Eq:DefSi}
\end{eqnarray}
In vacuum, the evolution equations consist of Eqs.\
(\ref{Eq:BSSN1}-\ref{Eq:BSSN8}) with $S=0$, $\hat{S}_{ij}=0$, $S^i =
0$. In order to obtain a solution to Einstein's vacuum field
equations, one also has to solve the constraints $H=0$, $M_i = 0$ and
$C_\Gamma^i = 0$. Below, we show that for $m=1$ it is sufficient to solve these
constraints on an initial Cauchy surface in the region {\em exterior}
to black holes. The constraint propagation system then guarantees that
these constraints hold at all events which are future to the initial surface and
outside the black hole regions, provided suitable
boundary conditions are specified at the outer boundary of the
computational domain.

%%%%%%%%%%%%%%%%%%%%%%%%%%%%%%%%%%%%%%%%%%%%%%%%%%%%%%%%%%%%%%%%%%%%%%%%%%%%%
\subsection{Hyperbolicity and characteristic speeds for the main system}
\label{SubSec:MainHypo}
%%%%%%%%%%%%%%%%%%%%%%%%%%%%%%%%%%%%%%%%%%%%%%%%%%%%%%%%%%%%%%%%%%%%%%%%%%%%%

The evolution system
(\ref{Eq:BSSN1}-\ref{Eq:BSSN8})
is first order in time and mixed first/second order in space. There
exist at least three different methods for analyzing hyperbolicity
(that is, the well--posedness of the Cauchy formulation) for such systems. The first
method consists in reducing the system to fully first order 
by introducing extra variables (and constraints) and to show that the
resulting first order system is strongly or symmetric hyperbolic (see
\cite{Reula98a} for definitions). The hyperbolicity of the BSSN
equations with a fixed shift and a densitized lapse or a Bona-Massó
type condition using this method has been established in Refs.\
\cite{Sarbach02a} and \cite{Beyer:2004sv}. The second method which was
developed in Refs.\ \cite{Kreiss:2001cu,Nagy:2004td} is also based on
a first order system. However, the reduction makes use of 
pseudo-differential operators. This has the advantage of not
introducing any new constraints. Unlike the first method, this results
in a unique first order system. The hyperbolicity of the BSSN
equations with a Bona-Massó type condition and a hyperbolic Gamma
driver type condition was shown in \cite{Beyer:2004sv} using this
method. Finally, the third method which has been proposed in Ref.\
\cite{Gundlach04a} and applied to BSSN in Ref.\ \cite{Gundlach:2004jp}
consists in finding an energy norm which, in the limit of frozen
coefficients, is conserved. This method has been shown
\cite{Gundlach:2005ta} to be equivalent to obtaining a first order
symmetric hyperbolic reduction with the first method.

Based on the second method, the following characteristic speeds with
respect to normal observers for the evolution system
(\ref{Eq:BSSN1}-\ref{Eq:BSSN8})
were found in \cite{Beyer:2004sv}: $0$, $\pm\mu_1$,
$\pm\mu_2$,$\pm\mu_3$, $\pm\mu_4$, $\pm\mu_5$ and $\pm\mu_6$ where
\begin{eqnarray}
\mu_1 = \sqrt{f}, \qquad
\mu_2 = \sqrt{\frac{4m-1}{3}}\; ,\qquad
\mu_3 = \sqrt{m},\\
\mu_4 = 1, \qquad
\mu_5 = \sqrt{GH}\; , \qquad
\mu_6 = \sqrt{\frac{4GH}{3}}\; .
\end{eqnarray}
When considering high-frequency perturbations of smooth solutions it
is possible to classify the characteristic fields as gauge fields,
constraint-violating fields and gravitational radiation
\cite{Sarbach02b,Calabrese02c}. According to this classification, the
fields propagating with speeds $\mu_1$, $\mu_5$ and $\mu_6$ correspond
to gauge modes, the fields propagating with speeds $\mu_2$ and $\mu_3$
to constraint-violating modes and the fields propagating with speeds
$\mu_4$ to gravitational radiation. As shown below, this statement can
be strengthened by noticing that $0$, $\mu_2$ and $\mu_3$ are the
characteristic speeds of the constraint propagation system. In fact,
it can be shown \cite{Reula:2004xd} under quite general assumptions
that the characteristic speeds of the constraint propagation system
are a subset of the speeds of the main evolution system.

In \cite{Beyer:2004sv} the following necessary conditions for strong
hyperbolicity are given: $f > 0$, $m > 1/4$ and $GH > 0$ or $f > 0 $,
$m > 1/4$ and $G=H=0$. (Notice that for $G=H=0$ the evolution equation
for the shift, Eq.\ (\ref{Eq:BSSN3}), decouples from the remaining
system.) If, in addition, the parameter $m$ and the functions $f$, $G$
and $H$ can be chosen such that the functions
\begin{displaymath}
\frac{4GH}{3f - 4GH}\; , \qquad
\frac{6(m-1)}{4m-1-4GH}\; , \qquad
\frac{2(m-1)GH}{m-GH}
\end{displaymath}
have smooth limits at $3f = 4GH$, $4m = 1 + GH$ and $m = GH$, then
strong hyperbolicity is guaranteed \cite{Beyer:2004sv}. For the
standard choice $m=1$ it is sufficient to verify that $f > 0$, $G H >
0$ and that the function $4GH/(3f - 4GH)$ has a smooth limit at $3f =
4GH$.

In the three-dimensional simulations below, we fix the functions $f$,
$K_0$ and $G$, $H$ and $\eta^i$ as follows. We choose the $1 + \log$
condition
\begin{displaymath}
f = \frac{2}{\alpha}\; ,\qquad
K_0 = 0,
\end{displaymath}
and the Gamma-driver shift condition
\begin{displaymath}
G = \frac{3}{4\alpha^2}\; , \qquad
H = e^{4\phi}, \qquad
\eta^i = \eta B^i
\end{displaymath}
with $\eta=1/2$. In this case, $\mu_1 = \sqrt{2/\alpha}$ and strong
hyperbolicity is guaranteed if the function $2\alpha e^{-4\phi} - 1$
does not cross zero. In our initial slices $\alpha \to 1$ and
$\phi \to 0$ in the asymptotic region while near black holes $\alpha >
0$ is small ($\alpha\approx0.3$ at the horizon) and $\phi$ positive.
Therefore there already exists a two-surface where the condition 
$2\alpha e^{-4\phi} - 1 \ne 0$ is violated in
the initial data. On the other hand, since this surface is a set of
zero measure in the computational domain there is hope that the
violation of our sufficient conditions at this surface might still
result in a well posed Cauchy problem. The numerical simulations below
show no apparent sign of instability.

%%%%%%%%%%%%%%%%%%%%%%%%%%%%%%%%%%%%%%%%%%%%%%%%%%%%%%%%%%%%%%%%%%%%%%%%%%%%%
\subsection{Hyperbolicity and characteristic speeds of the constraint
propagation system}
%%%%%%%%%%%%%%%%%%%%%%%%%%%%%%%%%%%%%%%%%%%%%%%%%%%%%%%%%%%%%%%%%%%%%%%%%%%%%

Next, we derive the constraint propagation system which describes the
propagation of constraint violations. We prove that for $1/4 < m \leq
1$ constraint violations inside a black hole region cannot propagate
to the outside.

A convenient way of finding the constraint propagation system is to
perform a $3+1$ decomposition of the contracted Bianchi identities,
$2\nabla^\mu R^{(4)}_{\mu\nu} - \nabla_\nu R^{(4)} = 0$, where one
sets the quantities $S$, $\hat{S}_{ij}$ and $S^i$ defined in Eqs.\
(\ref{Eq:DefS}-\ref{Eq:DefSi}) to zero. Taking into
account the definitions of the constraint variables $H$, $M_j$ and
$C^i_\Gamma$ defined in Eqs.\
(\ref{Eq:BSSNCons1}-\ref{Eq:BSSNCons3}) one finds
that they obey the linear evolution system 
\begin{eqnarray}
\hat{\partial}_0 H &=& -\frac{1}{\alpha}\, D^j(\alpha^2 M_j) 
-\alpha e^{-4\phi}\tilde{A}^{ij}\tilde{\gamma}_{ki}\partial_j C^k_\Gamma 
  + \frac{2\alpha}{3}\, K H,
\label{Eq:H}\\
\hat{\partial}_0 M_j
  &=& \frac{\alpha^3}{3} D_j( \alpha^{-2} H ) 
+ \alpha K M_j + M_i \partial_j\beta^i 
   + D^i\left( \alpha\left[ \tilde{\gamma}_{k(i}\partial_{j)}C^k_\Gamma 
\right]^{TF} \right),
\label{Eq:Mj}\\
\hat{\partial}_0 C^i_\Gamma &=& 2\alpha\, m\, \tilde{\gamma}^{ij} M_j\; .
\label{Eq:Ck}
\end{eqnarray}
In order to analyze this system, which is mixed first/second
order in space, we use the first method described in
Sec.\ \ref{SubSec:MainHypo} and reduce it to a first order symmetric
hyperbolic system. This allows us to establish the causal propagation
of the constraints via a standard energy inequality. Introducing the
additional constraint variable $Z_i{}^k = \partial_i C_\Gamma^k$,
$Z_{ij} = Z_i{}^k\tilde{\gamma}_{kj}$,  Eqs.\
(\ref{Eq:H}-\ref{Eq:Ck}) can be rewritten as the following
first order linear system: 
\begin{widetext}
\begin{eqnarray}
\hat{\partial}_0 H &=& -\frac{1}{\alpha}\, D^j(\alpha^2 M_j) 
 - \alpha e^{-4\phi}\tilde{A}^{ij} Z_{ij} + \frac{2\alpha}{3}\, K H,
\label{Eq:HFO}\\
\hat{\partial}_0 M_j &=& \frac{\alpha^3}{3} D_j( \alpha^{-2} H ) 
 + \alpha K M_j + M_i \partial_j\beta^i 
 + D^i\left( \alpha Z_{(ij)} \right)^{TF}
 - \sigma\alpha e^{-4\phi}
   \left[ \partial_k Z_j{}^k - \partial_j Z_k{}^k \right],
\label{Eq:MjFO}\\
\hat{\partial}_0 C^i_\Gamma &=& 2\alpha\, m\, \tilde{\gamma}^{ij} M_j\; .
\label{Eq:CkFO}\\
\hat{\partial}_0 Z_{ij} &=& 2m \partial_i\left( \alpha M_j \right)
 - 2\alpha m\tilde{\gamma}^{kl}(\partial_i\tilde{\gamma}_{jk}) M_l
 - 2\alpha\tilde{A}_{kj} Z_i{}^k
\nonumber\\
 &+& Z_{ik}\partial_j\beta^k + Z_{kj}\partial_i\beta^k
  + Z_i{}^k\tilde{\gamma}_{lj}\partial_k\beta^l
  - \frac{2}{3} Z_{ij} \partial_k\beta^k. 
\label{Eq:ZijFO}
\end{eqnarray}
\end{widetext}
Here, we have included in the right-hand side of Eq.\ (\ref{Eq:MjFO})
the term $\partial_k Z_j{}^k - \partial_j Z_k{}^k$ with an arbitrary
factor $\sigma$. Since $Z_i{}^k = \partial_i C_\Gamma^k$, this term is
identically zero. However, as we will see now, its addition allows
greater flexibility in obtaining a symmetric hyperbolic system. The
system (\ref{Eq:HFO}-\ref{Eq:ZijFO}) has the form
\begin{equation}
\hat{\partial}_0 C = \alpha
\left[ {\bf A}(u)^i\partial_i C + {\bf B}(u) C \right],
\label{Eq:ConstrEvol}
\end{equation}
where $C$ are the constraint variables, $u =
(\alpha,\beta^i,\phi,K,\tilde{\gamma}_{ij},\tilde{A}_{ij})$ are the
main variables, and ${\bf A}^i$, $i=1,2,3$, and ${\bf B}$ are
matrix-valued functions of $u$. Decomposing $Z_{ij} = \hat{Z}_{(ij)} +
Z_{[ij]} + \gamma_{ij} Z/3$ into its trace-free symmetric part,
$\hat{Z}_{(ij)}$, its antisymmetric part, $Z_{[ij]}$, and its trace,
$Z = \gamma^{ij} Z_{ij} = e^{-4\phi} Z_k{}^k$, and representing $C$ in
terms of the variables $C = (C_\Gamma^i,S_1 := 2m H + Z, S_2 := H +
2\sigma Z, M_j, \hat{Z}_{(ij)}, Z_{[ij]})$, the principal symbol ${\bf
A}({\bf n}) = {\bf A}(u)^i n_i$ is given by
\begin{equation}
{\bf A}({\bf n})\left( \begin{array}{c} 
 C_\Gamma^i \\ S_1 \\ S_2 \\ M_j \\ \hat{Z}_{(ij)} \\ Z_{[ij]})
\end{array} \right)
 = \left( \begin{array}{c}
 0 \\ 0 \\ (4m\sigma - 1) n^j M_j \\
 \frac{1}{3} n_j S_2 + (1-\sigma) n^i \hat{Z}_{(ij)} + \sigma n^i Z_{[ij]} \\
 2m  (n_{(i} M_{j)})^{TF} \\
 2m\, n_{[i} M_{j]}
\end{array} \right) .
\label{Eq:CPSymbol}
\end{equation}
Here $n^i \equiv \gamma^{ij} n_j$ and $n_i$ is normalized such that
$n_i n^i = 1$. This system is symmetric hyperbolic if and only if the
following inequalities hold:
\begin{displaymath}
4m\sigma - 1 > 0, \qquad
2m(1-\sigma) > 0, \qquad
2m\sigma > 0,
\end{displaymath}
which is equivalent to the two conditions $m > 1/4$ and $1/(4m) <
\sigma < 1$. Therefore, as long as $m > 1/4$ (which is also a
necessary condition for the {\em main} evolution system to be strongly
hyperbolic, see Sec.\ \ref{SubSec:MainHypo}) we can choose $\sigma$
between $1/(4m)$ and $1$ and obtain a symmetric hyperbolic
constraint propagation system. For the standard choice $m=1$, for
instance, we can choose $\sigma=1/2$ which is the case considered in
\cite{Brown2007b}. A symmetrizer ${\bf H} = {\bf H}^T$ is given by
\begin{eqnarray}
C^T {\bf H} C &=& \tilde{\gamma}_{ij} C_\Gamma^i C_\Gamma^j + S_1^2
 + \frac{1}{3(4m\sigma-1)}\, S_2^2 + \gamma^{ij} M_i M_j
\nonumber\\
 &+& \frac{1-\sigma}{2m}\,\gamma^{ik}\gamma^{jl}\hat{Z}_{(ij)}\hat{Z}_{(kl)}
  + \frac{\sigma}{2m}\,\gamma^{ik}\gamma^{jl} Z_{[ij]} Z_{[kl]}\; .
\nonumber
\end{eqnarray}
${\bf H}$ is positive definite and satisfies ${\bf H} {\bf A}({\bf n})
= {\bf A}({\bf n})^T {\bf H}$. The symmetrizer allows us to obtain an
energy-type estimate\footnote{Such estimates are a standard technique
in the theory of hyperbolic partial differential equations. In
particular, they allow one to prove uniqueness and continuous
dependence on the data and to establish the principle of finite
propagation speed. For references, see for instance \cite{Taylor96,
Reula:2004xd}.}  for the constraint variables $C$. For this, define
the four-current
\begin{displaymath}
J^\mu\partial_\mu = \frac{1}{2\alpha} C^T{\bf H} C\,\hat{\partial}_0
 - \frac{1}{2} C^T{\bf H}{\bf A}^i C\,\partial_i\; .
\end{displaymath}
By virtue of Eq.\ (\ref{Eq:ConstrEvol}) the current satisfies the
conservation law
\begin{equation}
\partial_\mu J^\mu \equiv \partial_t J^t + \partial_i J^i = C^T {\bf L} C,
\label{Eq:Conservation}
\end{equation}
with $2{\bf L} = {\bf H}{\bf B} + ({\bf H}{\bf B})^T +
\partial_t\left( \alpha^{-1} {\bf H} \right) - \partial_i\left( {\bf
H}{\bf A}^i + \alpha^{-1}\beta^i{\bf H}\right)$. Next, let $\Omega_T =
\bigcup\limits_{0 \leq t \leq T} \Sigma_t$ be a tubular region
obtained by piling up open subsets $\Sigma_t$ of $t=const$
hypersurfaces. This region is enclosed by the space-like hypersurfaces
$\Sigma_0$, $\Sigma_T$ and the surface ${\cal T} := \bigcup\limits_{0
\leq t \leq T} \partial\Sigma_t$, which is assumed to be
smooth. Integrating (\ref{Eq:Conservation}) over $\Omega_T$ and using
Gauss' theorem in $\mathbb{R}^4$ with the Euclidean metric, one
obtains
\begin{displaymath}
\int\limits_{\Sigma_T} J^t d^3 x = \int\limits_{\Sigma_0} J^t d^3 x
 + \int\limits_{\Omega} C^T {\bf L} C d^4 x
 - \int\limits_{\cal T} J^\mu e_\mu dS, 
\end{displaymath}
where $e_\mu$ is the outward unit one-form normal to ${\cal T}$ and
$dS$ the volume element on that surface. If the boundary term $J^\mu
e_\mu$ is positive or zero, one obtains the estimate
\begin{eqnarray}
\int\limits_{\Sigma_T} J^t d^3 x 
 &\leq& \int\limits_{\Sigma_0} J^t d^3 x
 + \int\limits_{\Omega} C^T {\bf L} C d^4 x
\nonumber\\
& = &\int\limits_{\Sigma_0} J^t d^3 x 
 + \int\limits_0^T \left( \int\limits_{\Sigma_t} C^T {\bf L} C d^3 x \right) dt
\nonumber\\
 &\leq& \int\limits_{\Sigma_0} J^t d^3 x 
 + b   \int\limits_0^T \left( \int\limits_{\Sigma_t} J^t d^3 x \right) dt,
\nonumber
\end{eqnarray}
where $b$ is a constant and where we have used the positivity of $J^t
= C^T{\bf H} C/(2\alpha)$ in the last step. By Gronwall's lemma, one
obtains the inequality
\begin{equation}
\int\limits_{\Sigma_t} J^t d^3 x 
 \leq e^{b t} \int\limits_{\Sigma_0} J^t d^3 x, \qquad
0 \leq t \leq T.
\label{Eq:ConstraintEnergyEstimate}
\end{equation}
Since $2\alpha J^t = C^T{\bf H} C$ is positive definite, this then
implies that $C = 0$ everywhere on $\Omega_T$ if $C = 0$ on $\Sigma_0$
which shows that it is sufficient to solve the constraints $C=0$ on
the initial slice $\Sigma_0$. In view of numerical applications,
however, the constraints are not exactly satisfied on
$\Sigma_0$. Instead, numerical errors introduced by solving the
constraint equations on a finite grid may be modeled by a sequence
$C_n$ of initial constraint violations which converges to zero as
the resolution goes to infinity. In this case, the estimate
(\ref{Eq:ConstraintEnergyEstimate}) shows that for each fixed $t\in
[0,T]$ the $L^2$-norm of the constraint variables $C_n$ converge to
zero on $\Sigma_t$.

In order to analyze the conditions under which the boundary term is
nonnegative it is convenient to expand the outward normal vector as
$e_\mu dx^\mu = N[a\,\alpha dt + n_i(dx^i + \beta^i dt)]$ where $n_i$
is normalized such that $\gamma^{ij} n_i n_j = 1$ and $N > 0$ is a
normalization factor. We set $N=1$ in the following since we are only
interested in the sign of the boundary term. Notice that $|a| < 1$ if
${\cal T}$ is time-like, $|a| > 1$ if ${\cal T}$ is space-like and
$|a|=1$ if ${\cal T}$ is a null surface. With this notation, the
boundary term is equal to
\begin{displaymath}
J^\mu e_\mu = \frac{1}{2} C^T 
\left[  a{\bf H} - {\bf H}{\bf A}^i n_i \right] C.
\end{displaymath}
Therefore, the condition for this boundary term to be positive or zero
is that all the eigenvalues of the matrix ${\bf A}^i n_i$ are smaller
than or equal to $a$. If ${\cal T}$ is a future event horizon, then
$a=1$ and this condition means that all of the eigenvalues must be
smaller than or equal to one. For the symbol given in Eq.\
(\ref{Eq:CPSymbol}) these eigenvalues are the characteristic speeds
(with respect to normal observers) and are
\begin{displaymath}
0, \qquad
\pm\mu_2 = \pm\sqrt{\frac{4m-1}{3}}\; ,\qquad
\pm\mu_3 = \pm\sqrt{m}\; .
\end{displaymath}
In particular, there are no superluminal speeds if $1/4 < m \leq 1$,
and in this case no constraint violations can propagate out of a black
hole. For the standard choice $m=1$ this condition is
satisfied. However, we also see that choosing $m > 1$ in the evolution
equation for $\tilde{\Gamma}^i$, Eq.\ (\ref{Eq:BSSN8}), yields
superluminal constraint speeds in which case constraint violations
inside a black hole can affect the exterior region.

Finally, we would like to point out that if the computational domain
contains time-like boundaries, then $|a| < 1$ and the sign of the
boundary term $J^\mu e_\mu$ is not automatically positive or zero. In
this case, boundary conditions need to be specified such that this
term is nonnegative and such that a well posed Cauchy problem is
obtained.

%%%%%%%%%%%%%%%%%%%%%%%%%%%%%%%%%%%%%%%%%%%%%%%%%%%%%%%
\subsection{Spherical symmetry}
%%%%%%%%%%%%%%%%%%%%%%%%%%%%%%%%%%%%%%%%%%%%%%%%%%%%%%%

The BSSN equations can be specialized to spherical symmetry as
described in Ref.~\cite{Brown:2007a}. The first step is to remove the
restriction $\tilde\gamma = 1$ on the determinant of the conformal
metric and replace it with an evolution equation for $\tilde\gamma$
\cite{Brown:2005aq}. In this paper we use the ``Eulerian evolution''
defined by $\partial_t \tilde\gamma = 2\tilde\gamma \tilde D_i
\beta^i$. The conformal connection functions are defined in terms of
the conformal Christoffel symbols by $\tilde\Gamma^i \equiv
\tilde\gamma^{jk} \tilde\Gamma^i_{jk}$.

The reduction to spherical symmetry is achieved by writing the
conformal metric as $ds^2 = \tilde\gamma_{rr} dr^2 +
\tilde\gamma_{\theta\theta} d^2\Omega$, where $d^2\Omega$ is the
metric for the unit two--sphere. The independent component of the
trace--free part of the extrinsic curvature is $\tilde A_{rr}$, and
the independent component of the conformal connection functions is
$\tilde\Gamma^r$.

In one dimension as in three, we use $1 + \log$ slicing and the
Gamma--driver shift condition (although here we use $\eta = 1$ for the
damping parameter).  For the main evolution system in spherical
symmetry the characteristic speeds are $0$, $\pm\mu_1$, $\pm\mu_2$ and
$\pm\mu_5$. Strong hyperbolicity is guaranteed for $f>0$, $m> 1/4$ and
$GH >0$ if the following conditions hold: $f \ne GH$ and $3GH + 1 \ne
4m$.

The constraint evolution system can be obtained by spherical reduction
of the system (\ref{Eq:H}-\ref{Eq:Ck}), or by direct calculation from
the 1D equations of motion \cite{Brown:2007a}.  Let $Z_r^r \equiv
\partial_r C^r_\Gamma$ and define the vector of constraints by $C
\equiv (H,M_r,C^r_\Gamma,Z^r_r)^T$. The constraint evolution equations
have the form $\hat{\partial}_0 C = \alpha[{\bf A}^r \partial_r C +
{\bf B} C]$ where ${\bf A}^r$ and ${\bf B}$ are functions of the BSSN
variables. The principal symbol is given by
\begin{equation}
   {\bf A}^r 
   = \left( \begin{array}{cccc} 0 & -2 e^{-4\phi}/{\tilde\gamma}_{rr} & 0 & 0 
		\\ 1/6 & 0 & 0 & 2 e^{-4\phi} /3 
		\\ 0 & 0 & 0 & 0 \\ 0 & 2m/{\tilde\gamma}_{rr} & 0 & 0 \end{array} \right) \ .
\end{equation}
The characteristic fields are $C^r_\Gamma$, $mH + e^{-4\phi} Z_r^r$,
and $H \pm 2\sqrt{3(4m-1)/{\tilde\gamma}_{rr}} e^{-2\phi} M_r +
4e^{-4\phi} Z^r_r$ with proper speeds $0$, $0$, and $\pm \mu_2 =
\pm\sqrt{(4m-1)/3}$, respectively. This system is symmetric hyperbolic
as long as $4m > 1$. For the case of primary interest, $m=1$, the
characteristic fields propagate along the normal and along the light
cone.

A symmetrizer can be constructed from the squares of the
characteristic fields:
\begin{widetext}
\begin{eqnarray}
     C^T {\bf H} C & = & (mH + e^{-4\phi}Z^r_r)^2 + (C^r_\Gamma)^2 
	+ (H + 2\sqrt{3(4m-1)/{\tilde\gamma}_{rr}} e^{-2\phi} M_r + 4e^{-4\phi} Z^r_r)^2 \nonumber\\
	& & + (H - 2\sqrt{3(4m-1)/{\tilde\gamma}_{rr}} e^{-2\phi} M_r + 4e^{-4\phi} Z^r_r)^2  \nonumber\\
	& = & (m^2 + 2)H^2 + (C^r_\Gamma)^2 + 24 (4m-1)e^{-4\phi} (M_r)^2/{\tilde\gamma}_{rr} 
	\nonumber\\  & & + 33 e^{-8\phi} (Z_r^r)^2
	 + 2e^{-4\phi}(m+8) H Z_r^r \ .
\end{eqnarray}
\end{widetext}
The spacetime current, defined by
\begin{equation}
	J^\mu \partial_\mu = \frac{1}{2\alpha} C^T {\bf H} C \,\hat{\partial}_0
	 -\frac{1}{2} C^T {\bf H} {\bf A}^r C \,\partial_r \ ,
\end{equation}
satisfies the conservation law
\begin{equation}
	\partial_\mu J^\mu \equiv \partial_t J^t + \partial_r J^r = C^T {\bf L} C 
\end{equation}
where $2{\bf L} = {\bf H}{\bf B} + {\bf B}^T {\bf H} + \partial_t
({\bf H}/\alpha) - \partial_r ({\bf H}({\bf A}^r + \beta^r/\alpha)
)$. As in the 3D case, we can now show that if $J^\mu e_\mu$ is
non--negative at the boundaries, then
\begin{equation}
	\int_{\Sigma_t} J^t \, dr  \leq e^{b t} \int_{\Sigma_0} J^t \, dr 
		\ ,\quad 0 \leq t \leq T 
\end{equation}
for some constant $b$. It follows that the constraints will vanish on
$\Sigma_t$ if they vanish on the initial hypersurface $\Sigma_0$.

The assumption that $J^\mu e_\mu $ is non--negative at the boundaries
for $m=1$ can be seen to hold at the black hole horizon by following
the same reasoning as in the three-dimensional case. We expand the
Euclidean normal one form $e_\mu$ as a linear combination of the unit
normal $u_\mu$ to the $t={\rm const}$ surfaces and the unit normal
$n_i$ to the two-dimensional boundary within the spacelike
hypersurfaces. As in the three-dimensional case, we have $e_\mu
dx^\mu = N[a\alpha dt + n_r (dx^r + \beta^r dt)]$ where $|a|=1$
characterizes a null surface.  By dropping the positive constant $N$,
we find $J^\mu e_\mu = C^T ( a {\bf H} - {\bf H} {\bf A}^r n_r ) C
/2$.  This shows that for a black hole horizon $J^\mu e_\mu$ is
positive if the eigenvalues of ${\bf A}^r n_r$ are less than or equal
to one. This is indeed the case for $m=1$, since the constraint
propagation system has eigenvalues $0$ and $\pm \sqrt{(4m-1)/3}$.

%%%%%%%%%%%%%%%%%%%%%%%%%%%%%%%%%%%%%%%%%%%%%%%%%%%%%%%%%%%%%%%%%%%%%%%%%%%%%
\section{Code descriptions}
\label{sec:code}
%%%%%%%%%%%%%%%%%%%%%%%%%%%%%%%%%%%%%%%%%%%%%%%%%%%%%%%%%%%%%%%%%%%%%%%%%%%%%

We use two different codes for the simulations presented in this
paper.  One of them is \texttt{McLachlan}, a three-dimensional
adaptive mesh refinement code, which uses the BSSN system of
equations, as described in \cite{Alcubierre99d, Alcubierre02a}, with
the gauge conditions described in section \ref{sec:formulation}
above. See section \ref{sec:formulation} for the exact form of these
equations.  The \texttt{McLachlan} code is a Cactus thorn which is
entirely generated by Kranc \cite{kranc04, Husa:2004ip, krancweb}
directly from equations and differencing stencils specified in
Mathematica format.  \texttt{McLachlan} uses the Cactus framework
\cite{Goodale02a, cactusweb1} and the Carpet mesh refinement driver
\cite{Schnetter-etal-03b, carpetweb}. The evolution code is fully
fourth order accurate in time and space.  We use fifth order spatial
interpolation at mesh refinement boundaries, using buffer zones as
described in \cite{Schnetter-etal-03b} to ensure stability and
convergence at mesh refinement boundaries, and using tapered grids as
described in \cite{Lehner:2005vc} to avoid the Berger--Oliger time
interpolation.  Our finite differencing operators are the standard
centered fourth order accurate first and second finite differencing
operators, except for the advection terms which are upwinded (also
fourth order).  We use a fourth order Runge--Kutta time integrator and
add fifth order Kreiss--Oliger dissipation \cite{Kreiss73} to the
right hand sides.  We apply standard radiation (Sommerfeld) boundary
conditions (as described in~\cite{Alcubierre02a}) to all components of
the evolved fields.  These boundary conditions are neither 4th order
convergent, nor constraint
preserving so non-convergent constraint violations will propagate inwards
from the outer boundary.  We therefore expect our code, in the limit of
infinite resolution, to be fully fourth order accurate only in the region that
is causally disconnected from the outer boundaries.  We place our
outer boundaries far enough out that they do not affect our wave
extraction procedure.

We center a stack of refined regions around the origin. Each region
is a cube with a resolution half of the next coarser region.

We also use a one-dimensional code to analyze several issues related
to the turducken technique in the context of a spherically symmetric
black hole. This is the code used in \cite{Brown:2007} and described
in detail as the ``Eulerian case" in \cite{Brown:2007a}.  The 1D code
uses a uniform radial grid with nodes at coordinate radii $r_j = (j -
1/2) \Delta r$, where $j = 1$, $2$, \emph{etc.} Fourth order finite
differencing is used for spatial derivatives, and fourth order
Runge--Kutta is used for the time update. No boundary conditions are
imposed at the origin for any of the variables except the shift vector
component $\beta^r$. That is, for all variables except $\beta^r$, the
finite difference stencil is shifted toward positive $r$ near the
origin so that no guard cells are needed. For $\beta^r$ we impose the
boundary condition $\partial_r \beta^r = 0$ at $r=0$ by using a fourth
order, one sided finite difference representation of $\partial_r
\beta^r$. The resulting guard cell value is
\begin{equation}\label{betaBC}
	\beta^r(0) = \frac{1}{22}\; \large[ 17\beta^r(1) + 9\beta^r(2)
        - 5\beta^r(3) + \beta^r(4) \large ]
\end{equation}
where the numbers in parentheses label grid points.  In the evolution
code, spatial derivatives of $\beta^r$ are computed by shifting the
finite difference stencil toward positive $r$ near the origin. In this
way only the single guard cell value $\beta^r(0)$ is needed. Numerical
experiments show that the condition (\ref{betaBC}) (or a similar one)
is needed for stability.

The one-dimensional code uses the variable $\chi \equiv e^{-4\phi}$
rather than $\phi$. The CFL (Courant-Friedrichs-Lewy) factor is $0.25$, and unless otherwise
stated the resolution for all the one-dimensional runs discussed here
is $\Delta r = M/200$.

%%%%%%%%%%%%%%%%%%%%%%%%%%%%%%%%%%%%%%%%%%%%%%%%%%%%%%%%%%%%%%%%%%%%%%%%%%%%%
\section{Single black hole evolutions and the end state}
\label{sec:KS}
%%%%%%%%%%%%%%%%%%%%%%%%%%%%%%%%%%%%%%%%%%%%%%%%%%%%%%%%%%%%%%%%%%%%%%%%%%%%%
In this section we investigate the effects of black hole smoothing on
the evolution of a spherically symmetric single black hole.

We start with the (Schwarzschild) isotropic black hole data
\begin{subequations}\label{Eq:oneDindat}
\begin{eqnarray}
	{\tilde\gamma}_{rr} & = & 1 \\
	{\tilde\gamma}_{\theta\theta} & = & r^2 \\
	e^{\phi} & = & 1 + M/(2r) \\
	\tilde\Gamma^r & = & -2/r 
\end{eqnarray}
\end{subequations}
along with $K = 0$ and $\tilde A_{rr} = 0$. For all one-dimensional
simulations, we use the $1+\log$ slicing and Gamma driver shift
conditions as described in Sec.~\ref{SubSec:MainHypo}.  The initial
values for the gauge variables are $\alpha = 1$ and $\beta^r = B^r =
0$.  Unless otherwise stated, all simulations use $m=1$.

The black hole interior is turduckened by making the replacement $r
\to \bar r(r)$ in the data (\ref{Eq:oneDindat}) for $0 \le r \le r_t$,
where
\begin{widetext}
\begin{equation}
	\bar r(r) \equiv  r_0 - (10r_0 - 4r_t) r^2/r_t^2 + (20r_0 - 6 r_t) r^3/r_t^3 
		- (15 r_0 - 4r_t) r^4/r_t^4 + (4r_0 - r_t)r^5/r_t^5 \ .
\end{equation}
\end{widetext}
The function $\bar r$ has the properties $\bar r (0) = r_0$, $\bar
r'(0) = 0$, $\bar r(r_t) = r_t$, $\bar r'(r_t) = 1$, $\bar r''(r_t) =
0$, and $\bar r'''(r_t) = 0$. Thus, with this turduckening, we extend
$r$ inside $r_t$ in such a way that the initial data are $C^3$ at
$r_t$ and nonsingular at $r=0$. The lack of smoothness at $r_t$
generates an error in any centered fourth order finite difference
derivative whose stencil extends across $r_t$. For first derivatives
the error is ${\cal O}(\Delta r^3)$, and for second derivatives the
error is ${\cal O}(\Delta r^2)$. Note that initially the horizon is
located at coordinate radius $r = M/2$. As long as $r_t < M/2$, the
turduckening lies entirely inside the black hole.

The turducken runs considered below use either $r_0 = 0.1M$, $r_t =
0.4M$ or $r_0 = 0.05M$, $r_t = 0.45M$. These two types of smoothing
are referred to as case TA and case TB, respectively. We compare these
results to results obtained with puncture data, denoted P, which is
equivalent to no turduckening (puncture evolution).

\subsection{Behavior of constraint violations}
Initially the constraint violations are restricted to the black hole
interior, and our analysis in Section~\ref{sec:formulation} shows that
for $m \leq 1$ they should stay there. Thus, we expect the constraints
to hold everywhere and at all times in the black hole exterior.

One obvious way to test this is to monitor the numerical constraints
as functions of time, and confirm that the violations introduced by
the turduckening do not leak out of the black hole.  This is indeed
the case. For smoothing types TA and TB, the initial Hamiltonian
constraint violation inside the black hole is $\sim 10^{-1}/M^2$,
while (at resolution $\Delta r = M/200$) the initial constraint
violation outside the black hole is $\sim 10^{-9}/M^2$. The constraint
violation outside the black hole remains $\sim 10^{-9}/M^2$ throughout
the evolution.

\begin{figure}[htbp]
	\subfigure[Time $t=3M$.]
	{\includegraphics[width=0.48\textwidth]{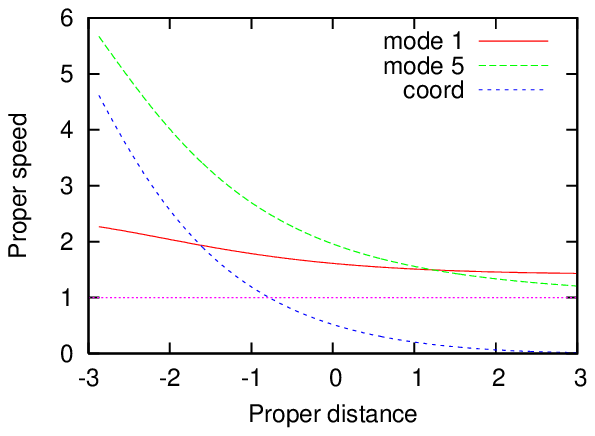}}
	\subfigure[Time $t=6M$.]
	{\includegraphics[width=0.48\textwidth]{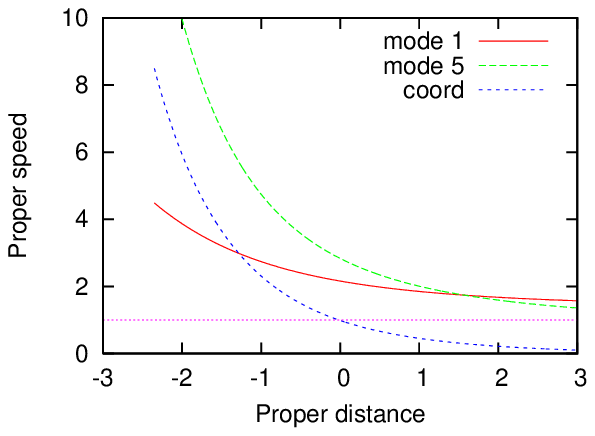}}
	\caption{Proper speeds as a function of proper distance from the 
	horizon for the turduckening case TA\@. The curve labeled 'coord' is the 
	proper speed of the coordinate system with respect to the normal observers, 
	$\sqrt{\tilde\gamma_{rr}} e^{2\phi} \beta^r/\alpha$. 
	The curve labeled ``mode 1'' is the proper speed $\mu_1 = \sqrt{2/\alpha}$. 
	The curve labeled ``mode 5'' is the proper speed $\mu_5 = \sqrt{3}e^{2\phi}/(2\alpha)$. 
	The horizontal line is light speed.}
	\label{fig:CoordSpeed}
\end{figure}

Moreover, we find that the region of constraint violation quickly
shrinks relative to the numerical grid, and the constraints quickly
loose memory of the turduckening. This comes about because the grid
points surrounding the origin acquire a radially outward velocity that
becomes superluminal within a time of a few $M$. The curves labeled
``coord'' in Figs.~\ref{fig:CoordSpeed} show the proper speed of the
coordinates in the radial direction with respect to observers at rest
in the spacelike hypersurfaces. The speed is plotted as a function of
proper distance from the black hole horizon, with the convention that
positive values are outside the black hole and negative values are
inside the black hole. What we see from these figures is that the
coordinate grid inside the black hole moves faster than the speed of
light ($c=1$) in the radially outward direction.  The region of
constraint violation moves causally, with speeds $0$ and $\pm
1$. Thus, the coordinates soon move outside the future light cone of
the stuffed region, and into the forward light cone of the initial
data that satisfies the constraints. The graphs in
Fig.~\ref{fig:CoordSpeed} were taken from simulations with smoothing
type TA\@. The graphs obtained with smoothing TB are nearly identical.

\begin{figure}
  \subfigure[Time $t=0M$. The horizon location is $r=0.5M$. At this scale the 
	curve P cannot be distinguished from the horizontal axis.]
  {\includegraphics[width=0.48\textwidth]{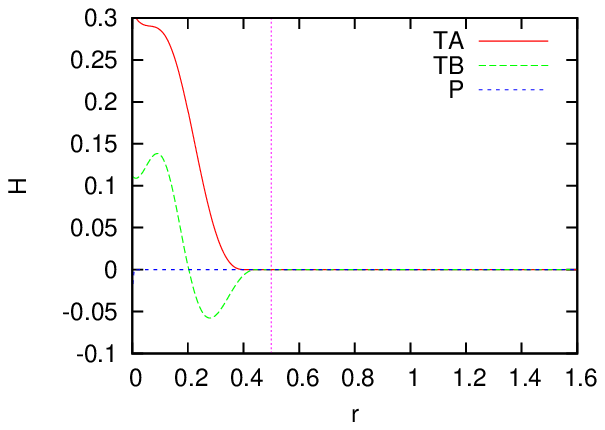}}
  \subfigure[Time $t=4M$. The horizon location is $r\approx 1.41M$.]
  {\includegraphics[width=0.48\textwidth]{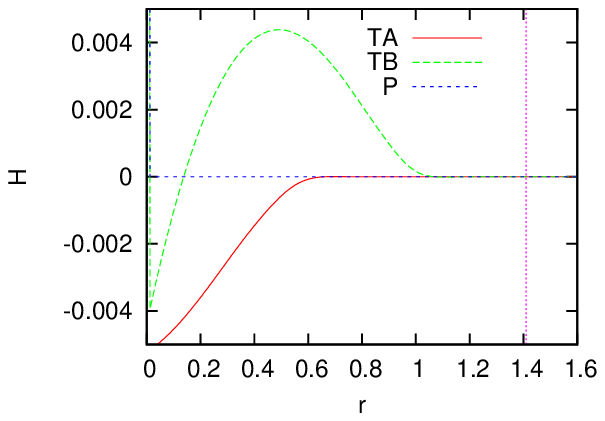}}
  \subfigure[Time $t=8M$. The horizon location is $r\approx 1.46M$. 
	At this scale the curves TA and P 
	cannot be distinguished from the horizontal axis.]
  {\includegraphics[width=0.48\textwidth]{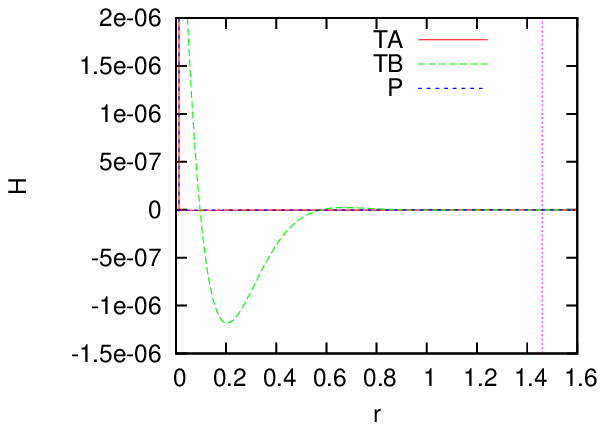}}
  \subfigure[Time $t=12M$. The horizon location is $r\approx 1.30M$.
	Curves TA and P are virtually identical. Curve TB shows 
	some small differences that disappear shortly after $t=12M$.]
  {\includegraphics[width=0.48\textwidth]{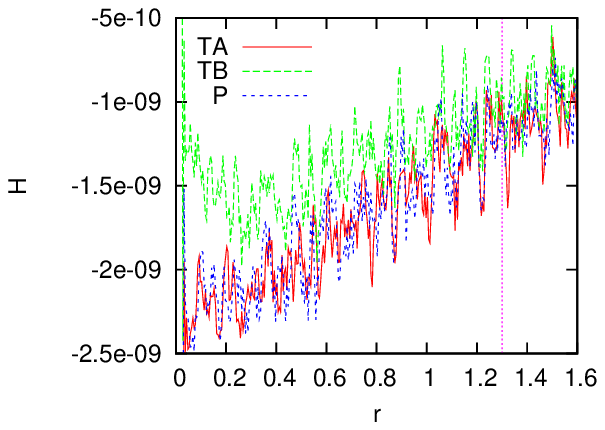}}
  \caption{Hamiltonian constraint versus coordinate radius $r$ at various 
	times, in units of $M$.  TA and TB denote two types of turduckening, and P is  
	puncture data. The vertical line shows the location of the black hole horizon.}
  \label{fig:Hdecay}
\end{figure}

In Figs.~\ref{fig:Hdecay} we plot the Hamiltonian constraint as a
function of coordinate radius for the initial data $t=0$ and for the
later times $t=4M$, $8M$, and $12M$. The data for turduckening types
TA and TB are shown, as well as for puncture data P\@. The constraint
violation does not propagate beyond the black hole horizon. The value
of the Hamiltonian constraint violation inside the black hole drops
from $\sim 0.1$ to $\sim 10^{-9}$ by $t=12M$. Beyond about $t=12M$,
the constraint data for TA and TB are everywhere (that is, also inside
the black hole) indistinguishable from each other, and
indistinguishable from the results obtained with puncture data P\@.

\begin{figure}
	\subfigure[Time $t=4M$. With $m=1.25$, the constraint violation is beginning to spread to the  black hole exterior. ]
	{\includegraphics[width=0.48\textwidth]{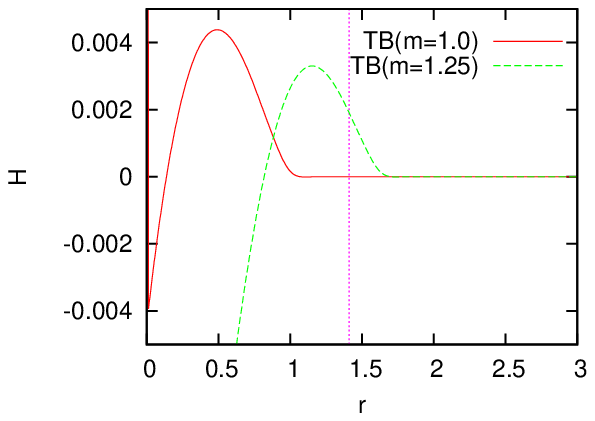}}
	\subfigure[Time $t=8M$. With $m=1.25$ a pulse of constraint violating data has moved beyond the black hole and continues 
	to propagate outward.]
	{\includegraphics[width=0.48\textwidth]{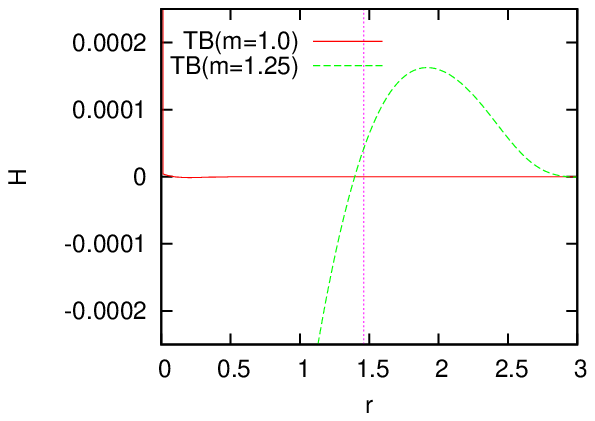}}
	\caption{Hamiltonian constraint versus coordinate radius $r$ for
	$m=1$ and $m=1.25$, where $m$ is the parameter that controls the mixing of the momentum constraint in the 
	equation of motion for $\Gamma^r$. The smoothing type TB is used in both cases. The vertical line shows the 
	location of the black hole horizon. As predicted by the theory, the
        constraint violations inside the black hole do propagate to the outside
        if $m>1$. }
	\label{fig:Hwithm}
\end{figure}

The success of the turduckening procedure depends on the constraint
violating modes propagating with speeds less than or equal to one. Our
analysis in Section~\ref{sec:formulation} predicts that this condition
is met for the BSSN family of evolution equations
(\ref{Eq:BSSN1})--(\ref{Eq:BSSN8}) if {\em and only if} $1/4 < m\leq
1$. In Fig.~\ref{fig:Hwithm} we plot the numerical Hamiltonian
constraint as a function of coordinate radius $r$ for $m = 1.25$. One
can clearly see that, as predicted by the theory, in this case the
constraint violation does propagate from the interior to the exterior
of the black hole. 
This illustrates the fact that for a given formulation of
the Einstein equations one cannot simply {\em assume} that the
constraints will propagate with non-superluminal speeds.

\subsection{Behavior of the coordinates}
\label{sec:coordinates}
The relatively large value of the radial component of the shift vector
moves the grid points beyond the region of constraint violation within
a time of a few $M$. However, the grid points do not move beyond the
influence of the turduckening.  Recall that the main evolution system
in spherical symmetry is strongly hyperbolic with characteristic
speeds $0$, $\pm \mu_1 = \pm \sqrt{2/\alpha}$, $\pm \mu_2 = \pm 1$,
and $\pm \mu_5 = \pm \sqrt{3} e^{2\phi}/(2\alpha)$. The modes with
speeds $\mu_1$ and $\mu_5$ can become superluminal. Figure
\ref{fig:CoordSpeed} shows the speeds $\mu_1$ and $\mu_5$ at times
$t=3M$ and $t=6M$, along with the speed of the coordinate grid
relative to the normal observers. Both speeds $\mu_1$ and $\mu_5$ are
larger than the coordinate speed at the black hole horizon (the origin
of proper distance). As discussed in Sec.~\ref{SubSec:MainHypo}, in
some sense the modes corresponding to $\mu_1$ and $\mu_5$ can be
associated with gauge freedom. We therefore expect that the
turduckening process {\it can} affect gauge conditions outside the
black hole. Although the full Einstein equations are satisfied in the
black hole exterior, independent of the smoothing, the slicing and
coordinate conditions outside the black hole can depend on the details
of the stuffing.

\begin{figure}[htbp]
	\subfigure[Time $t = 4M$.]
	{\includegraphics[width=0.48\textwidth]{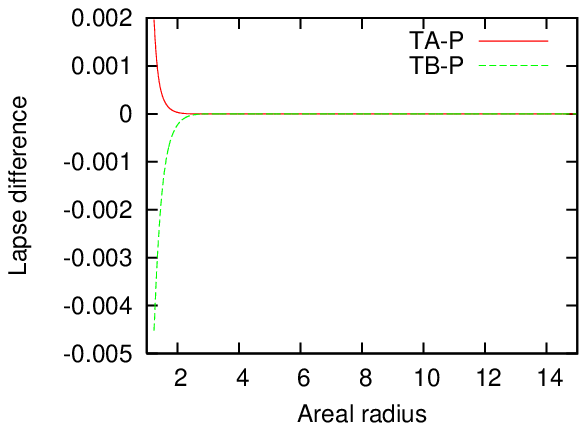}}
	\subfigure[Time $t= 8M$.]
	{\includegraphics[width=0.48\textwidth]{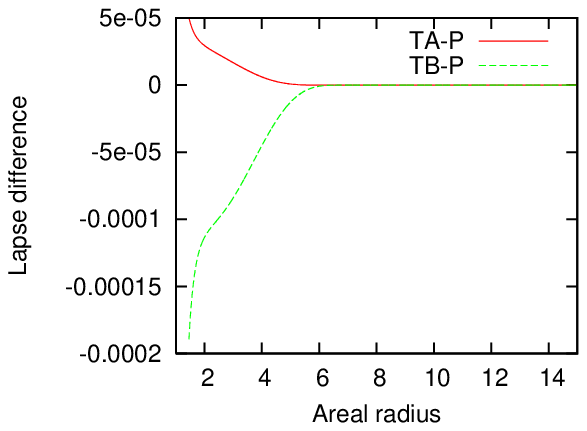}}
	\subfigure[Time $t=12M$.]
	{\includegraphics[width=0.48\textwidth]{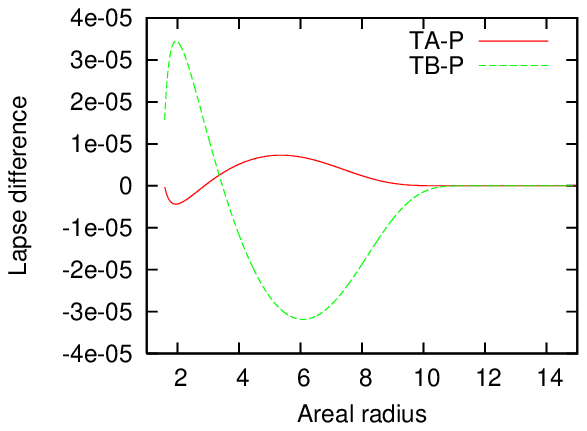}}
	\subfigure[Time $t=16M$.]
	{\includegraphics[width=0.48\textwidth]{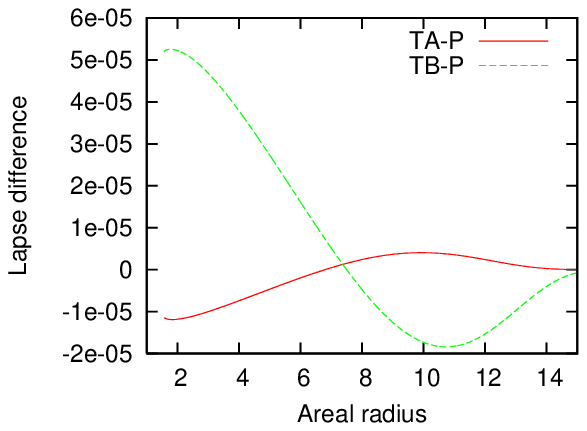}}
	\caption{The difference between lapse functions plotted versus 
	areal radius in units of $M$. The curve TA$-$P is the difference  
	for turducken data with stuffing TA and puncture data. The curve TB$-$P is the 
	difference for turducken data with stuffing TB and puncture data.}
	\label{fig:LapseDiff}
\end{figure}

Figure \ref{fig:LapseDiff} shows the differences between lapse
functions versus areal radius $R$, at four different times. For the
curve TA-P, we compute the difference between the lapse function for
smoothing TA and the lapse function obtained from puncture evolution
(no smoothing). These calculations require some explanation.  The
areal radius is not a monotonic function of the coordinate radius;
rather, the areal radius has a minimum at the black hole
throat. Therefore for each of the runs TA and P we use only the data
for which $R$ is an increasing function of coordinate radius.  We then
consider the overlap region in which $R$ is increasing for both data
sets TA and P\@. The difference in lapse values is computed by
interpolating this data onto a common grid that covers the overlap
region.  The differences TB$-$P are computed in the same way, but with
data from smoothing type TB\@.  Note that the overlap regions extend
inside the black hole horizon, which has areal radius $R=2M$.

The effect that the turduckening has on the slicing is relatively
small when compared to the nominal value of $1$ for the lapse, but is
clearly seen in Fig.~\ref{fig:LapseDiff}.  This effect begins inside
the black hole and propagates superluminally to the outside, where it
continues to spread radially outward.

\begin{figure}[htbp]
	\subfigure[Areal radius versus proper distance from the horizon at $t=M/4$.]
	{\includegraphics[width=0.48\textwidth]{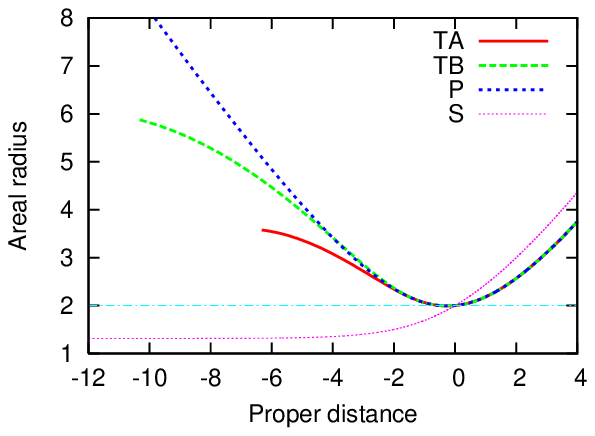}}
	\subfigure[Areal radius versus proper distance from the horizon at $t=50M$.]
	{\includegraphics[width=0.48\textwidth]{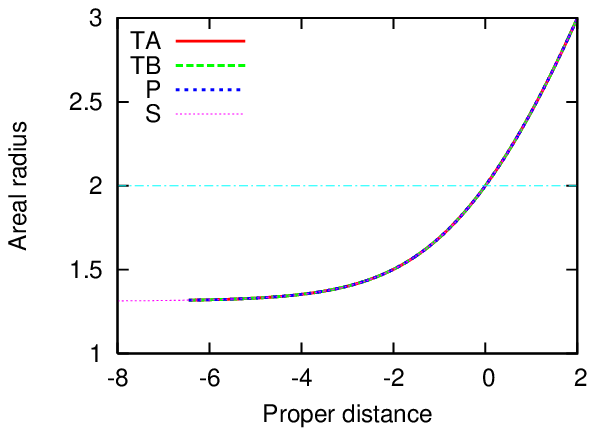}}
	\caption{Areal radius $R$ versus proper distance from the horizon. The horizontal 
	line at $R=2$ 
	is the horizon. By $t=50M$ the three curves (TA, TB, and P) are indistinguishable
	(at this scale) from the stationary $1+\log$ slice S\@.}
	\label{fig:endstate}
\end{figure}

The turduckening's effect on the slicing condition fades with time.
As the evolution proceeds, the data relax to a portion of a stationary
$1+\log$ slice of a Schwarzschild black hole, independent of the
initial stuffing details. This stationary $1+\log$ slice has a
``trumpet'' geometry \cite{Hannam2008a}. It is the same final slice
obtained with puncture evolution
\cite{Hannam:2006xw,Brown:2007,Baumgarte2007a}.  In
Fig.~\ref{fig:endstate} we graph the areal radius as a function of
proper distance from the horizon (with the convention that positive
distances are outside the black hole, negative distances are
inside). These plots show the data for smoothing types TA and TB, and
for no smoothing P, at early ($t=0.25M$) and late ($t=50M$) times. The
stationary $1+\log$ slice of Schwarzschild is shown as the curve
labeled S\@. Initially the $R$ versus proper distance relation shows a
strong dependence on smoothing. By $t=50M$ all of the data TA, TB, and
P have relaxed to approximate a portion of the trumpet slice S\@. At
this time, and with resolution $\Delta r = M/200$, the numerical
slices all end at proper distance $\approx -6.45M$.  Close inspection
of the data shows that near the end of the numerical slice, the areal
radius for the cases TA, TB, and P agree with one another to more than
seven decimal places; at proper distance $-6M$, we find $\alpha
\approx 1.32013$. The areal radius for a stationary $1+\log$ slice at
proper distance $-6M$ is $\alpha \approx 1.32018$.

\begin{figure}[htbp]
	\subfigure[Time $t=3M$, turducken radius $r_t = 0.2$.]
	{\includegraphics[width=0.48\textwidth]{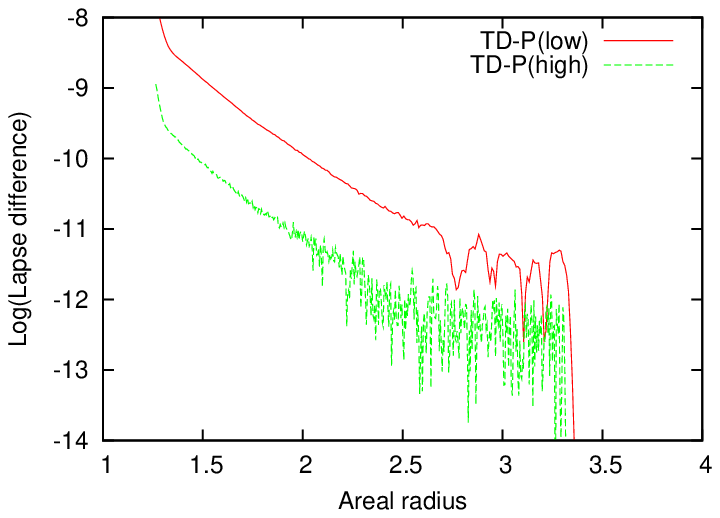}}
	\subfigure[Time $t=6M$, turducken radius $r_t = 0.2$.]
	{\includegraphics[width=0.48\textwidth]{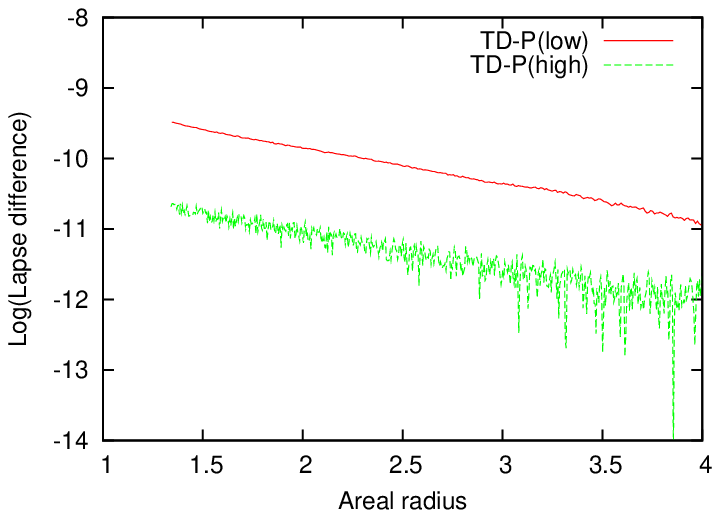}}
	\subfigure[Time $t=3M$, turducken radius $r_t = 0.3$.]
	{\includegraphics[width=0.48\textwidth]{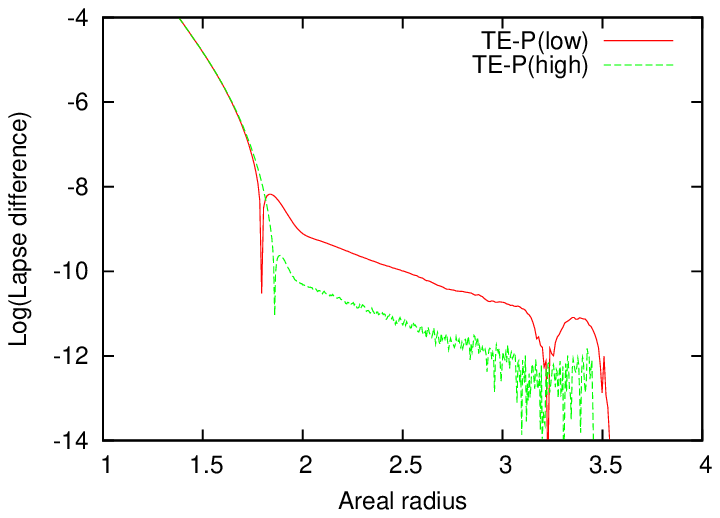}}
	\subfigure[Time $t=6M$, turducken radius $r_t = 0.3$.]
	{\includegraphics[width=0.48\textwidth]{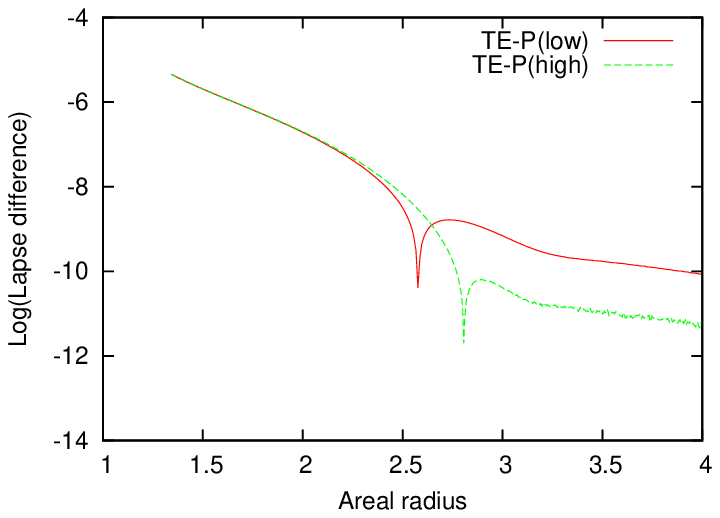}}
	\caption{Common logarithm of the difference between lapse functions. The top two subfigures 
	show the difference between case TD and puncture data for low 
	($\Delta r = M/100$) and high ($\Delta r = M/200$) 
	resolutions.  The bottom two subfigures show the difference 
	between case TE and puncture data for low and high resolutions. (Note that 
	the curves at time $t=3M$ appear to terminate at an areal radius of about $3.3$ 
	(case TD) or $3.5$ (case TE). This occurs 
	because, beyond these values, the lapse difference is exactly $0.0$ to 
	thirteen or more decimal places. The logarithm is 
	undefined for larger values of areal radius.) }
	\label{fig:lapsediffDE}
\end{figure}

Although the stuffing {\it can} affect the slicing beyond the black
hole horizon, it does not {\it always} do so. For the type of
simulations here considered, it appears that any stuffing that is
initially inside a coordinate radius of about $0.2M$ remains causally
disconnected from the black hole exterior.  In
Fig.~\ref{fig:lapsediffDE} we show the difference between lapse
functions for stuffing types TD, TE and puncture data P\@. The cases
TD and TE use turduckening radii $r_t = 0.2M$ and $r_t = 0.3M$,
respectively. Both cases use $r_0 = 0.02M$. The figure shows the
common logarithm of the lapse differences at $t = 3M$ and $t=6M$, for
low ($\Delta r = M/100$) and high ($\Delta r = M/200$) resolutions.

The top two subfigures show that the lapse difference TD$-$P converges
to zero with increasing resolution. On the log plots the difference
between low and high resolution curves is $\log(16) \approx 1.2$.
These results show that with its small turduckening radius, the data
for stuffing type TD are indistinguishable from puncture data in the black hole
exterior. The bottom two subfigures show the lapse difference
TE$-$P\@. In this case the difference converges to a nonzero value
which spreads from the black hole interior to its exterior.

Let us refer once again to Fig.~\ref{fig:CoordSpeed}, which shows the
gauge propagation speeds $\mu_1$, $\mu_5$ along with the coordinate
system speed. Observe that the speed $\mu_1 = \sqrt{f}$ depends on the
slicing condition (\ref{Eq:BSSN1}), while the speed $\mu_5 =
\sqrt{GH}$ depends on the coordinate shift conditions
(\ref{Eq:BSSN3}) and (\ref{Eq:BSSN4}). Thus it is the speed $\mu_1$ that
we examine closely here. Figure \ref{fig:CoordSpeed} shows that the
coordinate system moves faster than $\mu_1$ within a proper distance
of $\sim M$ of the puncture. This appears to be a common result,
independent of the stuffing details. If the stuffing initially extends
beyond this $\mu_1$--sphere, where the speed curve for $\mu_1$ crosses
the coordinate speed curve, then the stuffing's affect on the slicing
can propagate radially outward through the black hole horizon. This is
the case for data type TE\@.  If the stuffing is initially contained
entirely within the $\mu_1$--sphere, then the stuffing's affect on the
slicing is lost as the coordinate grid quickly moves beyond the
influence of mode $\mu_1$. This is the case for data type TD\@.

Note that the precise value of the turduckening radius where the
influence of the gauge speed $\mu_1$ is lost might depend on a number
of details of the simulation. For example, black hole spin, the
initial values of the lapse and shift, as well as the details of the
stuffing profile might affect whether or not the gauge modes will be
lost in the black hole interior.

\begin{figure}[htbp]
	\subfigure[Time $t=2M$.]
	%%{\includegraphics[width=0.48\textwidth]{oneDfigs/LapseDiffversusrTime3M}}
	{\includegraphics[width=0.48\textwidth]{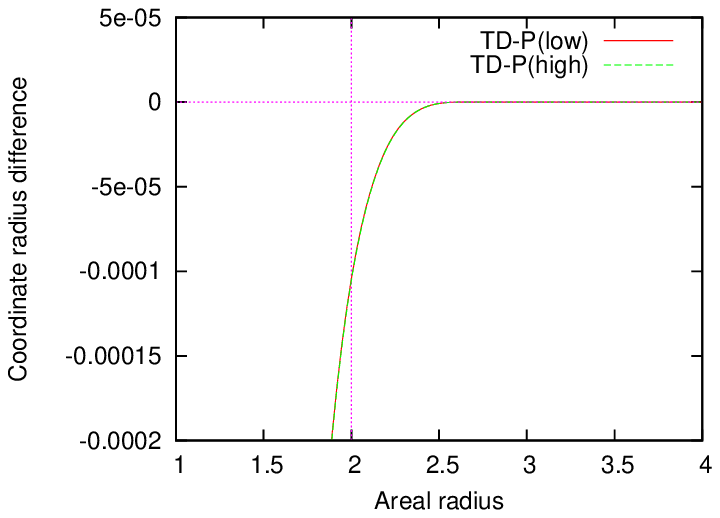}}
	\subfigure[Time $t=6M$.]
	%%{\includegraphics[width=0.48\textwidth]{oneDfigs/LapseDiffversusrTime6M}}
	{\includegraphics[width=0.48\textwidth]{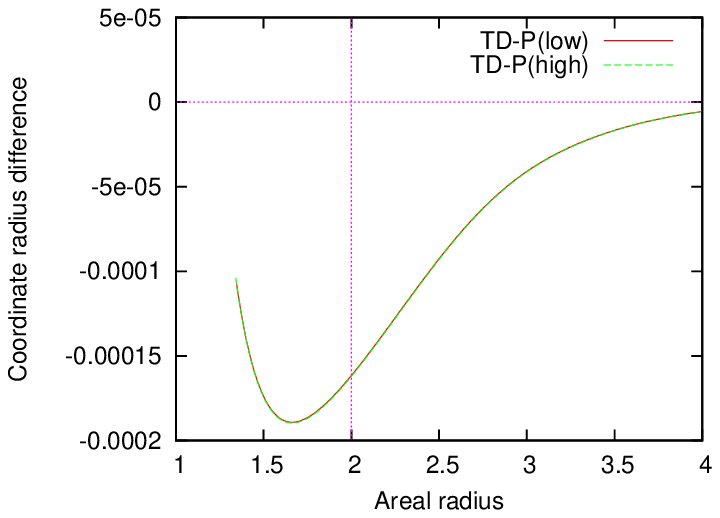}}
	\caption{The difference between coordinate radius for turducken data TD and 
	puncture data, as a function of areal radius. Both low 
	($\Delta r = M/100$) and high ($\Delta r = M/200$) 
	resolution results are shown, although the curves are difficult to distinguish. The 
	vertical line is the location of the black hole horizon. }
	\label{fig:rdiffversusR}
\end{figure}

Finally, let us observe that the gauge speed $\mu_5$ is larger than
the coordinate speed throughout the black hole interior.\footnote{This
observation partially explains why the one-dimensional code
\cite{Brown:2007a} seems to require boundary conditions at the origin
for the shift vector.}  Thus, we expect that the stuffing details will
affect the shift in the black hole exterior, regardless of how small
the stuffing region might be. This expectation is supported by the
results shown in Fig.~\ref{fig:rdiffversusR}. This figure shows the
difference in coordinate radius for turducken data TD and puncture data P 
as a function of areal radius. As
discussed above, in the limit of infinite resolution the slicing
outside the black hole appears to be identical for turducken data TD
and puncture data P\@.  Thus the difference shown in
Fig.~\ref{fig:rdiffversusR} is due to differences in the coordinate
grid. In particular, the radial shift in the coordinate grid in the
black hole exterior depends on the type of stuffing used.  At time
$t=2M$ the difference has just begun to cross the black hole
horizon. The difference continues to move into the
black hole exterior as the evolution proceeds.

%%%%%%%%%%%%%%%%%%%%%%%%%%%%%%%%%%%%%%%%%%%%%%%%%%%%%%%%%%%%%%%%%%%%%%%%%%%%%
\section{Three-dimensional black hole evolutions}
\label{sec:BY}
%%%%%%%%%%%%%%%%%%%%%%%%%%%%%%%%%%%%%%%%%%%%%%%%%%%%%%%%%%%%%%%%%%%%%%%%%%%%%

\subsection{Turduckening procedure in 3D}

Numerical initial data with black holes may contain singular
``puncture'' points or excised regions. If the data are incomplete due
to excision, the excised regions must be filled before the turducken
method can be used to evolve the data. Even if the data are complete,
as in the puncture case, it may be advantageous to replace the data
near singular points with data that is more smooth.

We experimented with various methods for turduckening the initial data
in the black hole interior.  We found that the details do not matter
much in practice, as long as the fields are sufficiently smooth across
the turduckening boundary and the spacetime remains unchanged within a
layer inside the horizon whose width is at least a few times the width
of the finite differencing stencil.  Empirically, a width of $10$ grid
points suffices for us; we expect this number to depend on the
particular differencing operators which are used.

One rather simple method for turduckening is blending, which fills the
excised region with arbitrary data, and then modifies some of the
non-excised grid points to create a smooth match. This has the
disadvantage that it may require quite a few non-excised grid points
inside the horizon---one needs to have sufficiently many grid points
to ensure the smooth match, plus the grid points which need to be left
unmodified.

Instead, we choose a method which leaves all given data unchanged and
fills in all excised points in a smooth manner. In particular, we
solve an elliptic equation of the form
\begin{equation}
  \label{eq:turducken}
  \left( \frac{\partial^n}{\partial x^n} +
         \frac{\partial^n}{\partial y^n} +
         \frac{\partial^n}{\partial z^n} \right) A = 0
\end{equation}
to fill the excised points of a quantity $A$, using standard centered
derivatives everywhere and using the given non-excised data as
boundary conditions.  Here $n$ is an even integer which controls the
smoothness of the resulting field $A$.  If $n=2$, $A$ will be only
$C^0$.  Increasing $n$ by 2 results in one additional derivative being
continuous, so that for $n=4$, $A$ is $C^1$, and for $n=6$, the
resulting $A$ is $C^2$.  Since we need to take two derivatives of the
metric, we choose $n=6$ to keep all derivatives continuous.  We will
show later that using $n=2$ still works, but leads to large errors.

Equation~(\ref{eq:turducken}) is linear, and we employ a standard
conjugate gradient method \cite{Shewchuk1994} to solve it.  This
numerical scheme is rather easy to implement, is robust, and it
converges reasonably quickly at the resolutions within reach on
current supercomputing hardware. The algorithm has been implemented in
the Cactus thorn \texttt{NoExcision} that we have made freely
available.\footnote{The Cactus thorn can be obtained via the command\\
\texttt{svn
checkout~https://svn.aei.mpg.de:/numrel/AEIThorns/NoExcision/trunk~NoExcision}}

Higher values of $n$ lead to smoother initial data.  It is also
possible to choose a non-zero right hand side in (\ref{eq:turducken}),
modifying the shape of the solution in the excised region.  In this
paper we do not take advantage of this additional freedom.  Other
turduckening procedures may also be possible and could be directly
integrated into the initial data solvers.

\subsection{Distorted rotating black hole evolutions}

We now turn to evolutions of single distorted, rotating black holes
and investigate the effects of black hole turduckening on the
numerical solution and the extracted waveforms.  We use single
puncture initial data with a Bowen-York extrinsic curvature
\cite{Ansorg:2004ds}.  We use a puncture mass $m_p = 0.751744$ and a
moderately large angular momentum parameter $S = 0.7$, resulting in a
black hole with irreducible mass $M_{\mathrm{irr}} \approx 0.925785$,
dimensionless spin $a/M = S/M^2 = 0.7$, and mass $M = 1.0$ (see e.g.\
Eq.~(27) in \cite{Dreyer02a} for a definition of the horizon mass used
here).  The ADM mass of the spacetime is $M_{\mathrm{ADM}} \approx 1.00252$.

We chose puncture initial data to be able to compare turducken and
puncture evolutions (see also Sec.\ \ref{sec:KS}).  We chose a
non-zero angular momentum to arrive at a more interesting,
non-spherically-symmetric case which also includes gravitational wave
emission, as the conformally flat rotating single punctures do not
represent the Kerr spacetime \cite{ValienteKroon:2003ux}.

We perform simulations of four different initial data setups.  The
first is a pure puncture setup where the turduckening procedure is not
applied.  In the other three cases, the puncture data are modified
inside of different turduckening radii: small ($r_t = 0.1M)$, medium
($r_t = 0.15M$), and large ($r_t = 0.2M$), but kept unchanged
everywhere else.  The radius of the initial apparent horizon is
$r_{\mathrm{hor}} = 0.3758 M$.  For all four cases we perform
simulations at three different resolutions: low ($dx = 0.024 M$),
medium ($dx = 0.016 M$) and high ($dx = 0.012 M$), as measured on the
finest refinement level.  Table~\ref{tab:npoints} lists the
approximate number of grid points between the horizon and the
turduckening region in each of these cases.
\begin{table}
  \begin{tabular}{lr@{\hspace{1em}}|@{\hspace{1em}}rrr}
           & Turduckening      & \multicolumn{3}{c}{Resolution}    \\
           & radius            & low       & med       & high      \\
           & $r_t$\hspace{1em} & $6\Delta$ & $4\Delta$ & $3\Delta$ \\\hline
    small  & $0.10 M$          &        11 &        17 &        23 \\
    medium & $0.15 M$          &         9 &        14 &        19 \\
    large  & $0.20 M$          &         7 &        11 &        15
  \end{tabular} 
  \caption{Number of grid points between the turduckening region
    and radius of the initial apparent horizon ($r_{hor} =
    0.3758 M$), for three turduckening radii $r_t$ and three grid
    resolutions $dx$.  The resolution $dx$ is given in multiples of
    $\Delta=0.04M$.  If there are too few grid points between the
    horizon and the turduckening region, information about the
    turduckening procedure escapes out of the black hole, which must
    be avoided.}
  \label{tab:npoints}
\end{table}
These runs were performed using the \texttt{McLachlan} code (see Sec.\
\ref{sec:code}), eight levels of mesh refinement, and the outer
boundaries located at $R=256M$.  Refinement boundaries were placed at
$R=[128, 64, 16, 8, 4, 2, 1]M$.  Fifth order Kreiss-Oliger dissipation
\cite{Kreiss73} was applied to all evolved variables.

We choose the initial lapse profile as
\begin{equation}
  \alpha_{\mathrm{init}} = \frac{1}{1+m_p/(2r)},
\end{equation}
which corresponds to the average of the isotropic lapse and unit
lapse.  Except for the puncture case, the lapse is further modified in
the turduckening region by smoothing it using
Eq.~(\ref{eq:turducken}) with $n=6$.  The shift and its time
derivative are initially set to zero.

\subsection{Numerical results}
\subsubsection{Constraints}
\begin{figure}
  \includegraphics[width=0.48\textwidth]{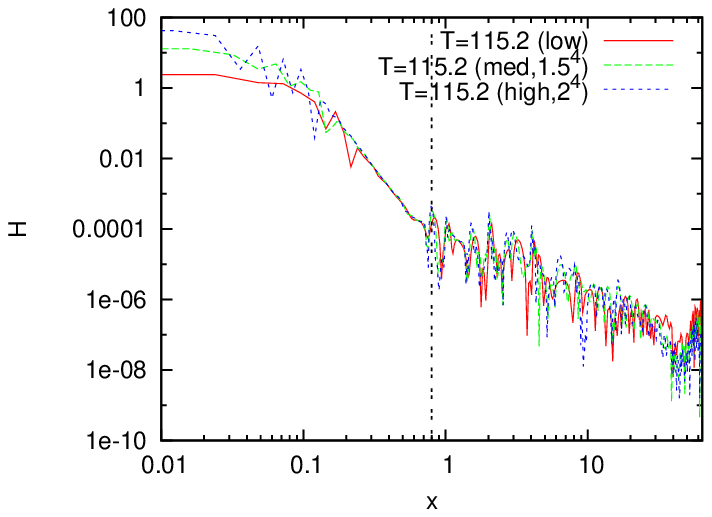}
  \includegraphics[width=0.48\textwidth]{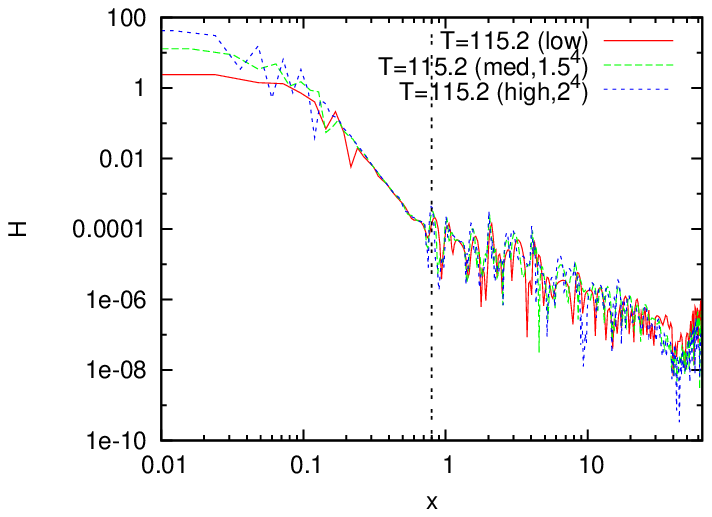}
  \caption{Convergence plot for the Hamiltonian constraint along the
    $x$ axis at $T=115.2M$ for the small turduckening region
    $r_t=0.1M$ (left plot) and for the puncture run (right plot).
    The medium and high resolution curves have been scaled
    for 4th order convergence. The vertical line shows the location
    of the horizon. The plots cover the range over which the waveforms
    were extracted (up to $R=60M$). At this time this region is still
    causally disconnected from the outer boundary.}
  \label{fig:hamiltonian}
\end{figure}

Figure~\ref{fig:hamiltonian} shows the convergence behaviour for the
Hamiltonian constraint for the small turduckening radius $r_t=0.1M$
and the puncture run along the $x$ axis at a late time $T=115.2M$ (any
late time could have been chosen, since the numerical spacetime has
essentially become stationary). The vertical line shows the location
of the horizon at this time. Similar results hold for the momentum
constraint.  These results show that at late times the constraints
converge to zero at a fourth order rate, except for a few grid points
near the origin.  Since the stationary solution we are approaching has
a $1/\sqrt{r}$ singularity in the conformal factor
\cite{Brown:2007,Hannam2008a}, this is to be expected, as a
neighborhood of the origin is under-resolved.
Note also that the convergence plots are practically identical for the
turducken and puncture simulations. These results are consistent with
the 1D results of Section~\ref{sec:KS}.

\begin{figure}
  \includegraphics[width=0.48\textwidth]{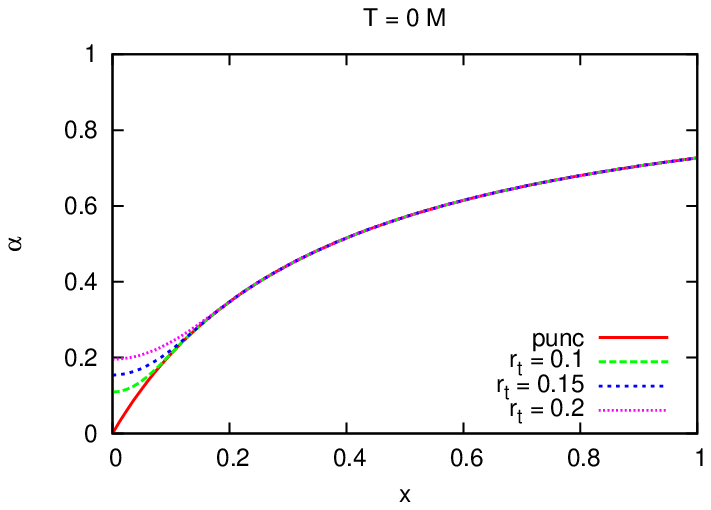}
  \includegraphics[width=0.48\textwidth]{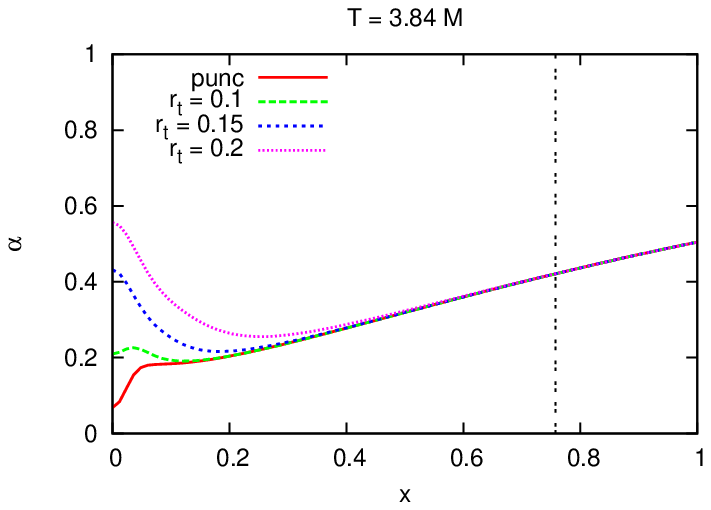}
  \includegraphics[width=0.48\textwidth]{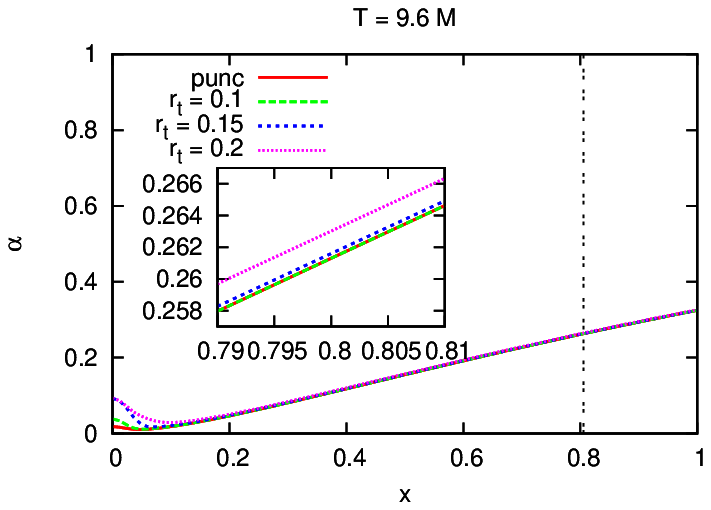}
  \includegraphics[width=0.48\textwidth]{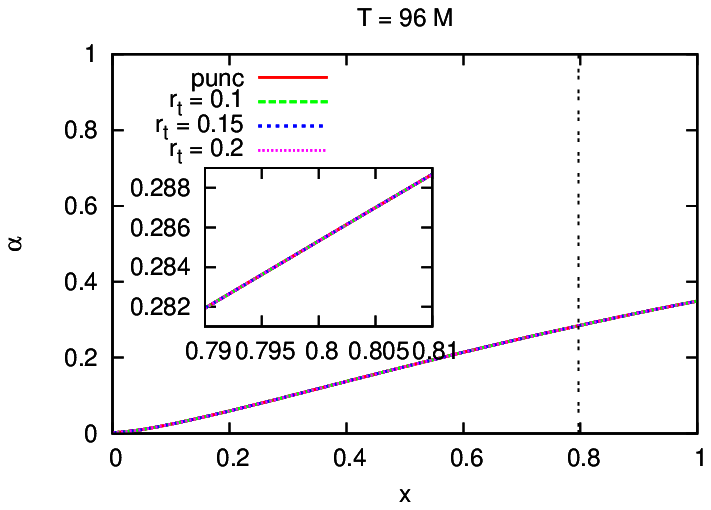}
  \caption{Lapse profile along the $x$ axis for the distorted single
    black hole evolutions at the times $T=0M$ (top left), $T=3.84M$
    (top right), $T=9.6M$ (bottom left), and $T=96M$ (bottom right)
    for the high resolution, comparing different turduckening radii.
    The vertical lines show the location of the apparent horizon.  The
    insets in the bottom graphs enlarge a small region near the
    horizon.  While the lapse profiles differ near the origin, they
    are very similar at the horizon, and their difference decreases
    with time.  However, for the medium and large turduckening radii,
    a superluminal gauge mode is clearly escaping. On the other hand
    the differences between the lapse profiles of the puncture and small
    turduckening radius cases converge to zero with resolution.}
  \label{fig:lapse_profile}
\end{figure}

\subsubsection{Lapse}
The behavior of the lapse is also consistent with the spherically
symmetric simulations of Section~\ref{sec:KS}.
Figure~\ref{fig:lapse_profile} shows the lapse profile along the $x$
axis at four different times ($T=0M$, $T=3.84M$, $T=9.6M$, and
$T=96M$) for the high resolution case.  The region with significant
differences in the slicing is at all times safely contained within the
horizon.  However, as can be seen from the inset in the bottom left
graph, there are real (but small) differences in the lapse function
also outside the horizon at early times.  These differences do not
converge away with resolution. In other words, the details of the
turduckening procedure result in real (but small) differences in the
slicing outside of the horizon.  With time, the differences become
smaller, and are no longer visible in the bottom right graph at
$T=96M$.

\subsubsection{Waveforms: convergence with resolution}
\begin{figure}
  \includegraphics[width=0.48\textwidth]{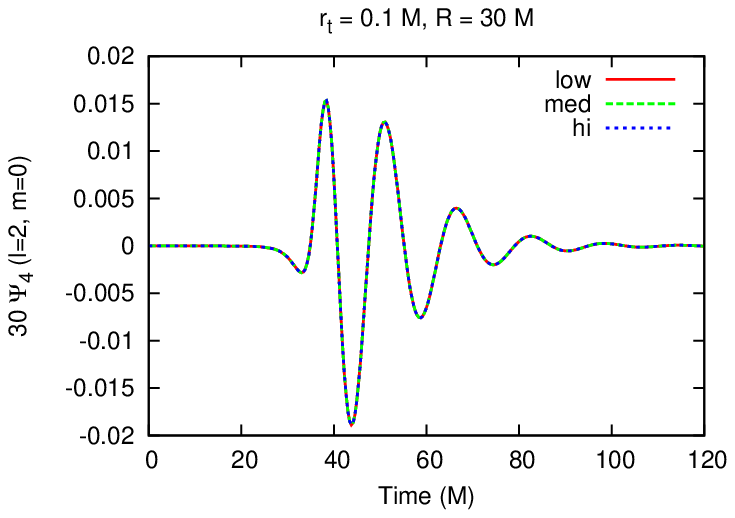}
  \includegraphics[width=0.48\textwidth]{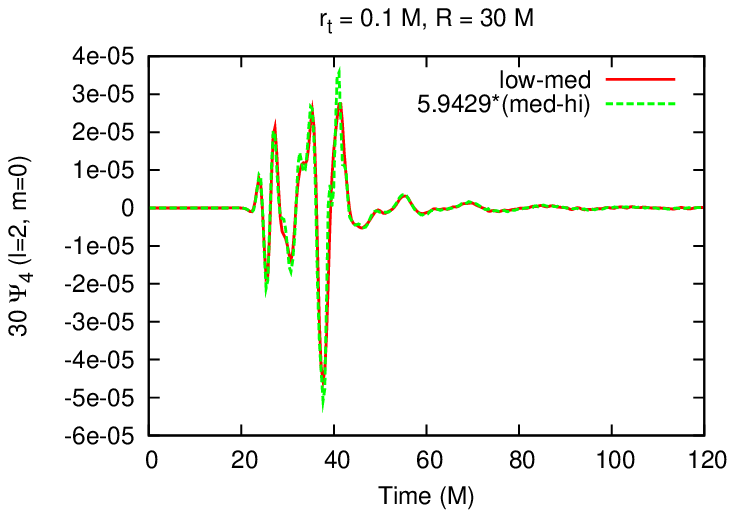}
  \includegraphics[width=0.48\textwidth]{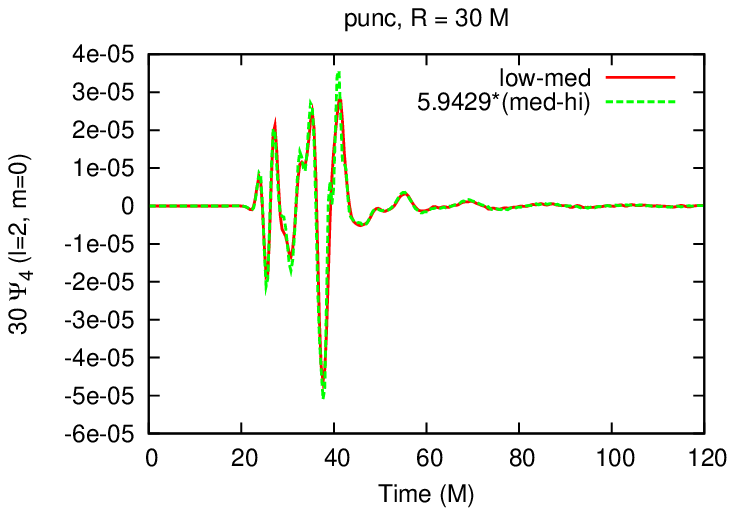}
  \includegraphics[width=0.48\textwidth]{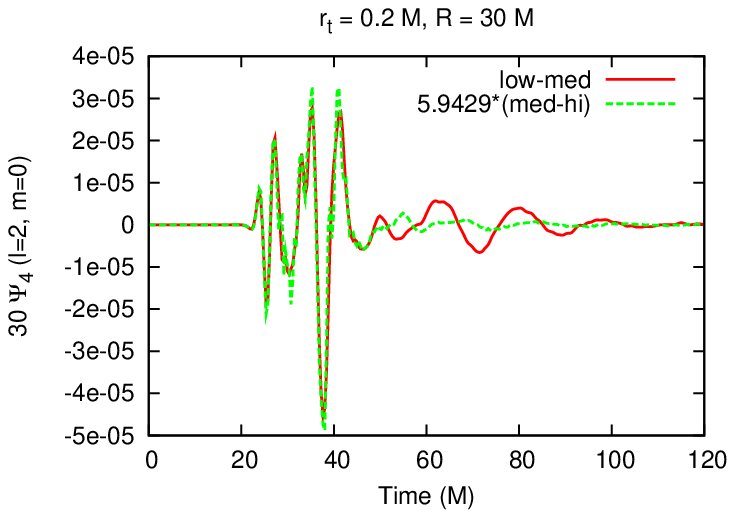}
  \caption{Effect of the turduckening radius on the $\ell=2, m=0$ mode
    of gravitational waveforms $r \Psi_4$, extracted at $R=30M$.  The
    top left graph compares all three resolutions, and the top right
    graph shows their differences, scaled for 4th order convergence,
    both for the small turduckening region $r_t=0.1M$.  The bottom
    left graph shows the scaled differences for a pure puncture run,
    and the fact that it looks virtually identical indicates that
    $r_t=0.1M$ is a good choice for these resolution.  On the other
    hand, the bottom right graph shows the same for the large
    turduckening radius $r=0.2M$ where noticeable differences are visible.
    We attribute this to the smaller number of grid points between the
    turduckening region and the initial apparent horizon (see table
    \ref{tab:npoints}), so this case is not yet in the
    convergent regime.}
  \label{fig:waveform30}
\end{figure}

We extract the $\ell = 2, m=0$ mode of the Weyl scalar $\Psi_4$ on
coordinate spheres at four different radii ($R=30M, R=40M, R=50M$, and
$R=60M$), choosing the commonly used hypersurface-adapted tetrad
described e.g.\ in \cite{Pollney:2007rve}.
Figure~\ref{fig:waveform30}
shows convergence tests for the waveforms obtained with different
turducken radii and with puncture evolutions, in all cases extracted at
$R=30M$. 

The top left graph shows the waveform with small turducken radius $r_t
= 0.1M$ at three different resolutions. The top right graph shows the
differences between the different resolutions, scaled for fourth order
convergence.  The fact that the curves look so similar is a
manifestation of clean fourth order convergence.

The bottom left graph shows the scaled differences for a pure puncture
run, also indicating clean fourth order convergence.  The case
$r_t=0.15M$ (not shown here) behaves similarly.

However, the scaled differences for $r_t=0.2M$ (bottom right graph)
looks noticeably different.  We attribute this to the smaller number
of grid points between the turduckening region and the initial
apparent horizon (see table \ref{tab:npoints}), so that, in the low
resolution case, there are too few gridpoints between the turduckening
region and the horizon for the numerical waveforms to be effectively
isolated from the stuffing.  The curve for the difference between the
medium and high resolution waveforms is very similar to the other
cases, which indicates that it is only the low resolution run that is
not in the convergent regime.  We expect, though we have not yet
confirmed this with numerical experiments, that a convergence test
with a higher minimal resolution will also show a lack of convergence
if the turduckening region is chosen so that $r_t$ is only $\sim 7$
(or fewer) gridpoints away from the horizon at the minimal resolution.
Thus, as a rule of thumb, we expect that one should always choose the
turduckening region so that $r_t$ is more than $\sim 7$ gridpoints
away from the horizon.  Finite differencing schemes that differ from
the one used in our code would probably require different
limits. Experimentation would be necessary on a case--by--case basis.

\subsubsection{Waveforms: comparison among different solution methods}
\begin{figure}
  \includegraphics[width=0.48\textwidth]{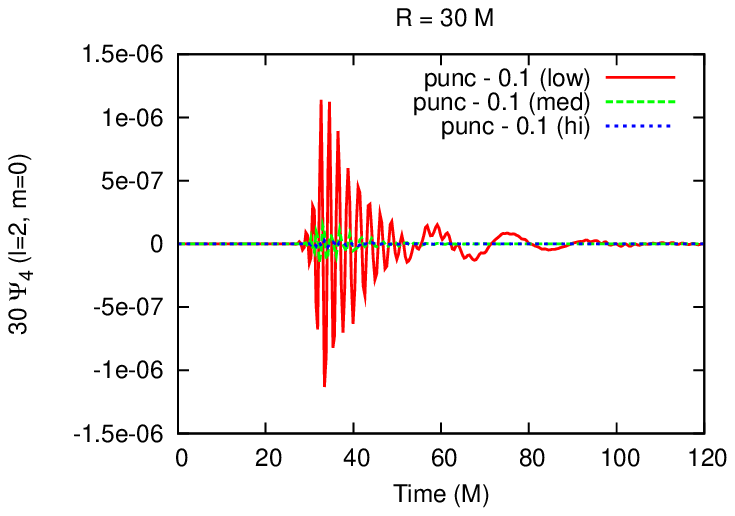}
  \includegraphics[width=0.48\textwidth]{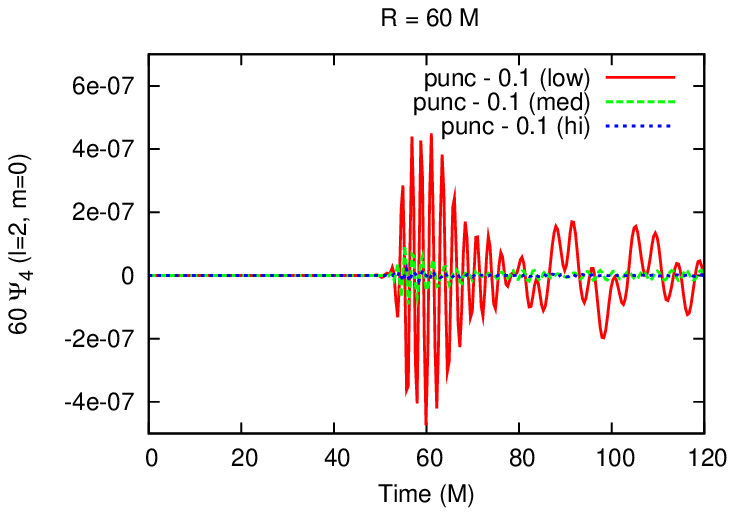}
  \includegraphics[width=0.48\textwidth]{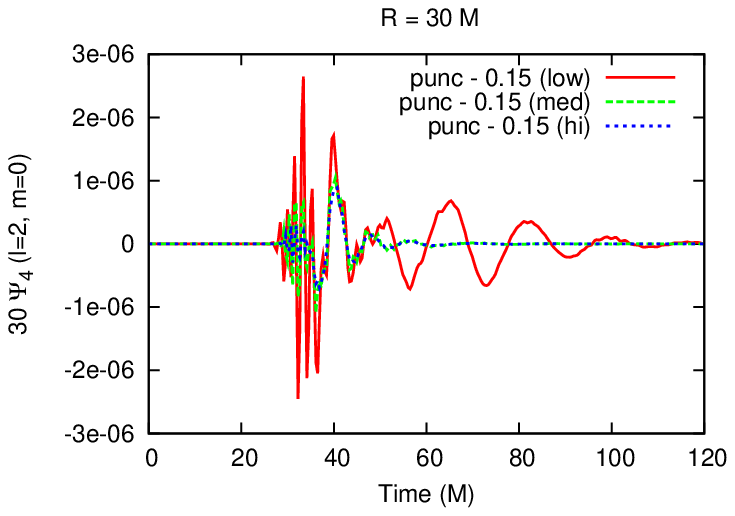}
  \includegraphics[width=0.48\textwidth]{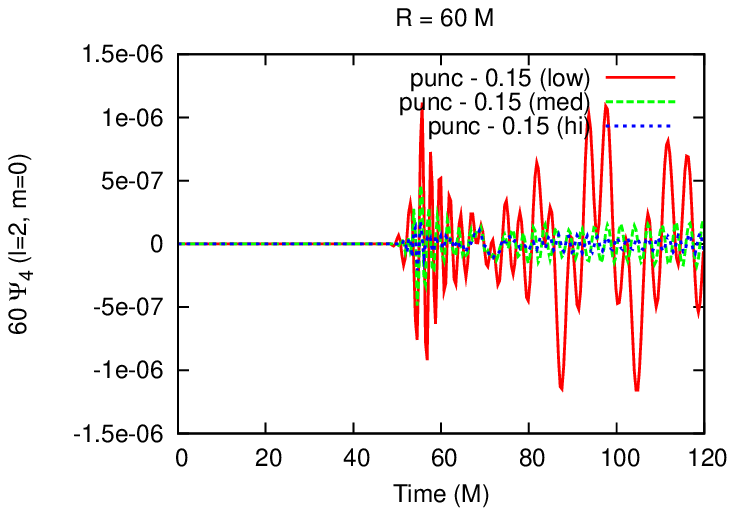}
  \includegraphics[width=0.48\textwidth]{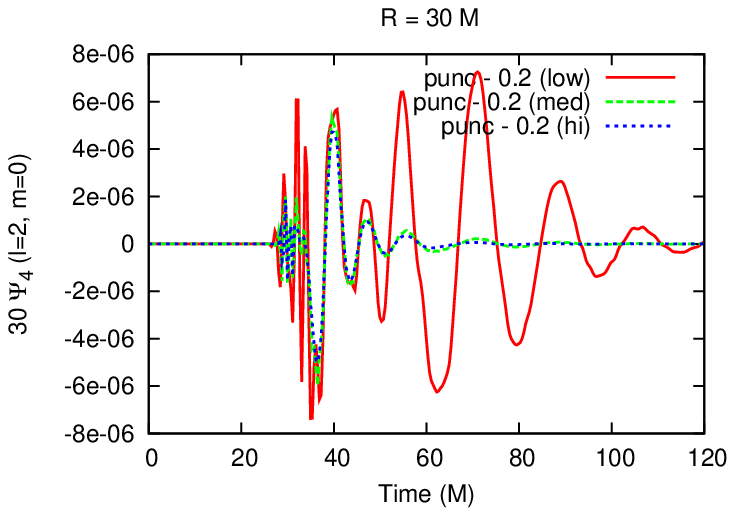}
  \includegraphics[width=0.48\textwidth]{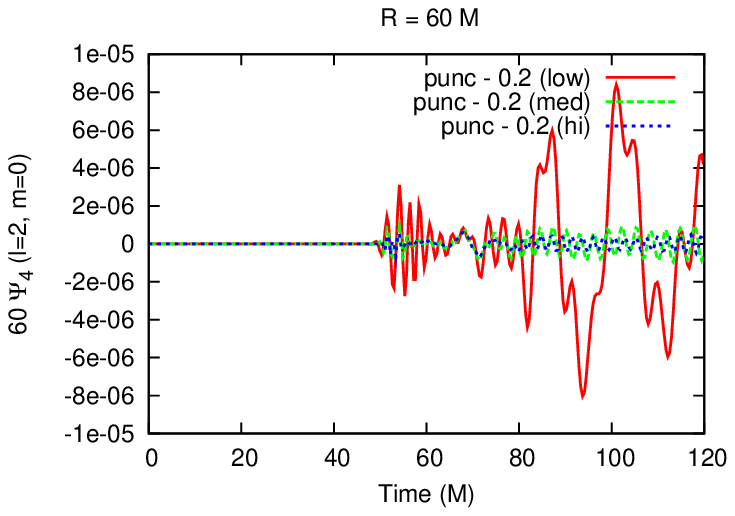}
  \caption{Difference between the $\ell=2, m=0$ mode of $r \Psi_4$ for
    a puncture evolution and for three different turduckening radii,
    extracted at $R=30M$ (left-hand graphs) and $R=60M$ (right-hand
    graphs).  See the main text for the discussion.}
  \label{fig:waveform-turducken-30}
\end{figure}

Figure~\ref{fig:waveform-turducken-30} shows, for the three resolutions, the differences between
the puncture waveform and the turducken waveforms with 
turduckening radii $r_t=0.1M$, $r_t=0.15M$, and $r_t=0.2M$. In all cases
two extraction radii are used: $R=30M$ and $R=60M$.
As can be seen from the figure, the difference between the puncture
waveform and the $r_t=0.1$ one goes to zero as the resolution is
increased. This is not true for the $r_t=0.15$ and $r_t=0.2$ cases,
where the difference, especially clear at $R=30M$, does not converge
to zero with increasing resolution.  Differences in the slicing do
appear to affect the extracted gravitational wave signal.  However, at
the larger extraction radius $R=60M$, the non--convergent part of the
waveform difference is significantly smaller. 

As already discussed, there are gauge modes that travel at
superluminal speeds with our parameter choices and differences in
gauge are able to propagate beyond the horizon.  This is true independently
of whether the initial data are pure puncture or turduckened data.
The fact that we only
find differences between the puncture waveforms and turducken ones
with larger $r_t$ does not mean that we consider the puncture
waveforms and the turducken waveforms with small $r_t$ to be correct,
and any deviation from them to be an error.  As described above, we
modify the
initial lapse profile away from the puncture profile only inside of the
turduckening region. We would expect that a pure puncture run with a different
initial lapse profile (for example, constant lapse equal to one) 
would also yield significant differences in the lapse at the location of the
detector.  Instead, the conclusion
is that since wave extraction is carried out at a finite radius, our
resolutions are high enough so that the seemingly small differences in
slicing lead to noticeable differences in waveforms. Furthermore, even
though this is a rotating black hole, which cannot be compared
directly to a Schwarzschild black hole, we believe that this is the
same effect as seen in the 1D spherically symmetric case (see
Fig.~\ref{fig:lapsediffDE} and its discussion in
section~\ref{sec:coordinates}). Namely, differences in slicing get
trapped inside the black hole for sufficiently small turducken radius
(while the puncture evolution can be considered as the limit
$r_t\rightarrow 0$). Note also that the slicing differences (and consequently
the waveform differences) are transient since both puncture and turducken runs
approach the same trumpet slice at late times.

What we have seen is that, through a proper convergence test, we can
detect differences in the waveforms due to different slicings at
fixed, finite extraction radii. However, it is far from clear that
these differences are of any practical importance. For example, for
each of our turduckening radii, the largest difference (at the highest
resolution) between a turducken waveform and the puncture one is
$6\times10^{-6}$. The amplitude of the wave is about $0.015$, so the
relative difference is about $4\times 10^{-4}$.

\begin{figure}
  \includegraphics[width=0.48\textwidth]{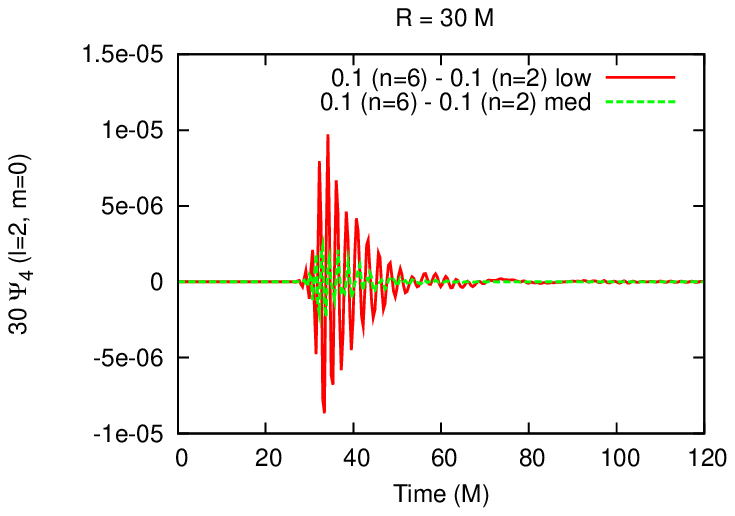}
  \includegraphics[width=0.48\textwidth]{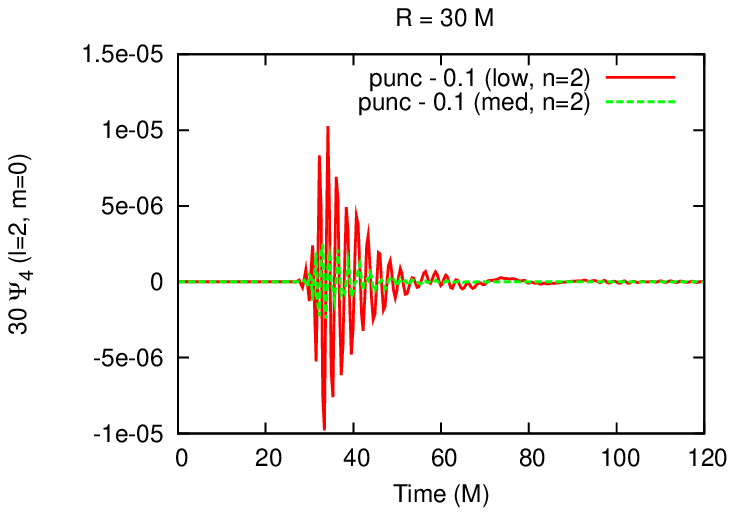}
  \caption{Difference between the $\ell=2, m=0$ mode of $r \Psi_4$ for
    the turducken runs with $n=6$ and $n=2$ (both with $r_t=0.1$) extracted
    at $R=30M$ (left plot). The right plots shows the same for the puncture
    run and the turducken with  $r_t=0.1$ and $n=2$.}
  \label{fig:waveform-turducken-2nd-30}
\end{figure}

Finally, we performed experiments at low ($dx=0.024M$) and medium
($dx=0.016M$) resolutions with a turduckening radius $r_t =0.1$, but
with $n=2$ in equation~(\ref{eq:turducken}), so that the evolution
fields are only $C^0$ at the boundary of the turduckening region.  In
Figure~\ref{fig:waveform-turducken-2nd-30} the left graph shows the
difference between the waveforms from runs with $n=2$ and $n=6$ while
the right graph shows the difference between the $n=2$ and puncture
waveforms. These two graphs are almost identical and a comparison with
the top left graph in Figure~\ref{fig:waveform-turducken-30} explains
why.
As can be seen, the difference between the puncture waveforms and the
$r_t=0.1$ waveforms is about 10 times larger for $n=2$ than $n=6$.
Note that in the $r_t=0.1M$ case the turduckening region is far enough
inside the black hole that we do not see gauge differences, so this
must be caused by increased numerical noise coming from the less
smooth data in the $n=2$ case.

%%%%%%%%%%%%%%%%%%%%%%%%%%%%%%%%%%%%%%%%%%%%%%%%%%%%%%%%%%%%%%%%%%%%%%%%%%%%%
\section{Final remarks}
\label{sec:Final}
%%%%%%%%%%%%%%%%%%%%%%%%%%%%%%%%%%%%%%%%%%%%%%%%%%%%%%%%%%%%%%%%%%%%%%%%%%%%%

In this paper we have analyzed in detail several aspects of the
turduckening technique for evolving black holes.

First we presented a detailed analytical study of the constraints
propagation for a rather general family of BSSN-type formulations of
the Einstein equations. We could appropriately identify a sub-family
for which the constraints propagate within the light cone and give a
rigorous justification of the turduckening procedure. At the same
time, we showed that in other subfamilies the constraint violations do
move superluminally. As a consequence, in those cases, smoothing the
interior of black holes will result in constraint violations that
propagate to the outside.

Through high-resolution spherically symmetric numerical simulations we
analyzed in detail the behavior of the constraints and gauge modes. In
particular, we confirmed to very high accuracy the predictions of our
analytical study. We found that the numerical constraints are
preserved outside the black hole when the theory predicts so, and are
violated otherwise. We also found cases in which gauge modes do
propagate superluminally, escaping from the black hole, and cases in
which they are trapped in the inside. These differences in gauge modes
have consequences for wave extraction, as discussed below.  We also
observed that, when the constraints are guaranteed to propagate within
the light cone, the region of constraint violations inside the black
hole shrinks with time, and that the same final stationary
configuration seems to be approached, regardless of the details of the
turduckening procedure. We also provided explanations for these
features.

Finally, we presented detailed three-dimensional simulations of single
distorted black holes, comparing turduckened and puncture
evolutions. We studied the effect that these different methods have on
the coordinate conditions, constraint violations, and extracted
gravitational waves.  We found the waves to agree up to small but
non-vanishing differences. Our convergence tests showed that those
differences are not numerical artifacts but true features of the
solution, caused by superluminal gauge modes escaping from the black
hole. We also found that these differences in waveforms decay with
increasing extraction radius.

%%%%%%%%%%%%%%%%%%%%%%%%%%%%%%%%%%%%%%%%%%%%%%%%%%%%%%%%%%%%%%%%%%%%
\begin{acknowledgments}
%%%%%%%%%%%%%%%%%%%%%%%%%%%%%%%%%%%%%%%%%%%%%%%%%%%%%%%%%%%%%%%%%%%%

O.S. wishes to thank Dar\a'{\i}o Nu\~nez for help in deriving the
characteristic speeds of the constraint propagation system, P.D. and
E.S. thank Christian D. Ott and Jian Tao for help with the
\texttt{McLachlan} BSSN code. We also wish to thank Ian Hawke and
Denis Pollney for many discussions which ultimately led to the
turduckening procedure. Our numerical calculations used the Cactus
framework \cite{Goodale02a, cactusweb1} with a number of locally
developed thorns, J. Thornburg's apparent horizon finder
\cite{Thornburg95, Thornburg2003:AH-finding}, the GNU Scientific
Library \cite{gslweb}, and the LAPACK \cite{lapackweb} and BLAS
\cite{blasweb} libraries from the Netlib Repository \cite{netlibweb}.
This research was supported in part by NSF PIF Grant 0701566
\emph{XiRel} \cite{xirelweb}, and NSF SDCI Grant No. 0721915 \emph{Alpaca}
\cite{alpacaweb} to Louisiana State University, NSF Grant No. PHY-0600402
to North Carolina State University, Grant CIC 4.19 to Universidad
Michoacana, PROMEP UMICH-PTC-195 from SEP Mexico, CONACyT Grant No.
61173, NSF Grant No. 0801213 to the University of Maryland, and the
TeraGrid allocation TG-MCA02N014 to Louisiana State University. We used the
supercomputing resources Peyote at the AEI, Santaka at LSU, Eric,
Queen Bee, and Tezpur at LONI, and Abe and Tungsten at the NCSA\@. We
also employed the resources of the Center for Computation \&
Technology at Louisiana State University, which is supported by
funding from the Louisiana legislature's Information Technology
Initiative.

\end{acknowledgments}

\bibliographystyle{bibtex/apsrev-titles}

\bibliography{bibtex/references}

\end{document}